\documentclass[aps,pre,reprint,superscriptaddress]{revtex4-2}

\usepackage[T1]{fontenc}
\usepackage[utf8]{inputenc}

\usepackage{amsmath,amssymb,amsfonts,amsthm,mathtools,bm}
\usepackage{graphicx,xcolor}
\usepackage{float}
\usepackage[colorlinks=true,linkcolor=blue,citecolor=blue,urlcolor=blue]{hyper%
ref}
\DeclareRobustCommand{\redtext}[1]{{\color{black}\hypersetup{linkcolor=blue,ci%
tecolor=blue,urlcolor=blue}#1}}

\theoremstyle{definition}
\newtheorem*{thm}{Theorem}%
\newtheorem*{defi*}{Definition}
\newtheorem*{prop}{\redtext{Proposition}}%

\newcommand{\dd}{\mathrm{d}}

\newcommand{\kB}{k_\mathrm{B}}

\allowdisplaybreaks[1]

\newcommand{\parhead}[1]{\section{#1}}

\begin{document}

\title{Geometric Decomposition of Entropy Production Based on General
\(f\)-Divergences}

\author{Rihito Nagase}
\affiliation{Department of Applied Physics, The University of Tokyo, 7-3-1
Hongo, Bunkyo-ku, Tokyo 113-8656, Japan}

\author{Shoki Sugimoto}
\affiliation{Department of Applied Physics, The University of Tokyo, 7-3-1
Hongo, Bunkyo-ku, Tokyo 113-8656, Japan}
\affiliation{\redtext{Nonequilibrium Quantum Statistical Mechanics RIKEN
Hakubi Research Team, RIKEN Pioneering Research Institute (PRI), Wako, Saitama
351-0198, Japan}}

\author{Takahiro Sagawa}
\affiliation{Department of Applied Physics, The University of Tokyo, 7-3-1
Hongo, Bunkyo-ku, Tokyo 113-8656, Japan}
\affiliation{Quantum-Phase Electronics Center (QPEC), The University of Tokyo,
7-3-1 Hongo, Bunkyo-ku, Tokyo 113-8656, Japan}
\affiliation{Inamori Research Institute for Science (InaRIS), Kyoto-shi, Kyoto
600-8411, Japan}

\date{\today}

\begin{abstract}
We construct a general class of excess--housekeeping decompositions of entropy
production based on $f$-divergences for overdamped Langevin systems.
The excess entropy production rate equals the relaxation rate of the
$f$-divergence toward a relaxation target and also admits an optimal-transport
representation, while the housekeeping velocity preserves the same relaxation
target.
Our framework recovers the Hatano--Sasa and Maes--Neto{\v{c}}n{\'y}
decompositions as special cases, and unifies their relaxational and
optimal-transport aspects within a common structure.
Moreover, our framework yields physically meaningful decompositions beyond
these two by taking a desired distribution as the relaxation target.
\end{abstract}

\maketitle

\parhead{Introduction}\label{sec:intro}
Stochastic thermodynamics quantifies energy cost and irreversibility
\redtext{in small systems coupled to heat
baths}~\cite{Sekimoto2010,Jarzynski2011,Seifert2012,
VanDenBroeckEsposito2015,Ciliberto2017}.
The entropy production is a central quantity in stochastic thermodynamics that
quantifies the irreversibility of a process~\cite{Seifert2012}.
\redtext{To distinguish the origins of irreversibility, entropy production
rate (EPR) is decomposed into an excess part associated and a housekeeping
part~\cite{OonoPaniconi1998,HatanoSasa2001,Ge2009,
VanDenBroeckEsposito2010FP,SpeckSeifert2005,GeQian2010,
EspositoVanDenBroeck2010Detailed,EspositoVanDenBroeck2010ME,
SpinneyFord2012,MaesNetocny2014,DechantSasaIto2022PRE,
YoshimuraEtAl2023,DechantSasaIto2022PRR,TrepagnierEtAl2004}.}
This excess--housekeeping decomposition yields various thermodynamic insights,
such as a geometric interpretation of
dissipation~\cite{ItoDechant2020,DechantSasaIto2022PRR,Ito2018,NakazatoIto2021%
},
finite-time thermodynamic
bounds~\cite{AurellEtAl2011,AurellEtAl2012,SchmiedlSeifert2007,SivakCrooks2012,
ShiraishiFunoSaito2018,FalascoEsposito2020,VanVuHasegawa2021},
and dissipation estimates from observed
data~\cite{RoldanParrondo2010,GingrichRotskoffHorowitz2017,LiEtAl2019,OtsuboEt%
Al2020,
ManikandanEtAl2020,KimEtAl2020,SkinnerDunkel2021}.
\par
Two representative formulations for overdamped Langevin systems are the
Hatano--Sasa (HS) decomposition~\cite{HatanoSasa2001} (also known as the
adiabatic--nonadiabatic decomposition~\cite{VanDenBroeckEsposito2010FP}) and
the Maes--Neto{\v{c}}n{\'y} (MN) decomposition~\cite{MaesNetocny2014}.
The HS decomposition separates EPR into an excess part which describes
relaxation toward the instantaneous steady state and a housekeeping part
required to maintain the nonequilibrium steady current.
Its excess EPR coincides with the relaxation rate of the Kullback--Leibler
(KL) divergence relative to the instantaneous steady
distribution~\cite{VanDenBroeckEsposito2010FP}.
\redtext{On the other hand,} the MN decomposition assigns the cost of the
distribution-changing part to the excess part and the rest to the housekeeping
part.
The excess EPR coincides with the minimal cost required to realize the given
\redtext{evolution of the probability distribution}, formulated as an optimal
transport problem~\cite{BenamouBrenier2000,DechantSasaIto2022PRR,NakazatoIto20%
21,AurellEtAl2011,Dechant2022,VanVuSaito2023}.
Note that, for Markov jump systems and deterministic chemical reaction
networks, counterparts of the HS
decomposition~\cite{EspositoVanDenBroeck2010Detailed,EspositoVanDenBroeck2010M%
E,RaoEsposito2018,RaoEsposito2016} and MN
decomposition~\cite{Kobayashi2022HessianDecomposition,YoshimuraEtAl2023,Kolchi%
nsky2026GeneralizedFreeEnergy} have also been developed.
\par
Mathematically, {\redtext{both} the HS decomposition and the MN decomposition}
have been \redtext{formulated} as orthogonal decompositions of the velocity
field with respect to an
appropriate inner product~\cite{DechantSasaIto2022PRR,DechantSasaIto2022PRE,
YoshimuraEtAl2023,Ito2024}.
However, there is still room for unifying, within a consistent structure, the
important properties discussed in the previous paragraph, namely, the
relaxational representation of the HS excess EPR and the optimal-transport
representation of the MN excess EPR.\par 
Another challenge is to find decompositions \redtext{other than the HS and MN
decompositions}.
For overdamped Langevin systems, some approaches based on
three-part~\cite{DechantSasaIto2022PRE} and
nonorthogonal~\cite{Sireci2026ConservativeDissipation} decompositions have
been proposed.
However, beyond HS and MN, a systematic decomposition of EPR into two
nonnegative parts based on orthogonal velocities \redtext{has} yet to be
established.
\par
In this study, we introduce a class of excess--housekeeping decompositions
based on general $f$-divergences, which unifies HS and MN and yields further
systematic decompositions.
\redtext{For a consistently chosen $f$-divergence and relaxation target
$\gamma_t$, we decompose} the velocity $\bm v_t$ into an excess part given by
the gradient-flow velocity of the $f$-divergence relative to an instantaneous
relaxation target $\gamma_t$, and a housekeeping part that preserves
\redtext{the same $\gamma_t$}.
This decomposition is orthogonal and thus yields an excess--housekeeping
decomposition of EPR.
\par
In our construction, the excess EPR equals the relaxation rate of the
$f$-divergence.
It also admits an optimal-transport representation as the minimal squared norm
among velocities \redtext{that, if applied to $\gamma_t$, would induce the
same instantaneous change as $v_t$ would.}
These properties generalize those of the HS and MN decompositions,
respectively.
Our construction thus integrates these two decompositions under the common
structure in which the housekeeping part preserves the relaxation target
$\gamma_t$.\par
Specifically, the HS and MN decompositions arise as two special cases, which
provides a mathematical basis for why these two decompositions have {the
above-mentioned representative} properties.
\redtext{In addition, our framework allows us to quantify the entropic cost of
approaching a desired target state, through an appropriate choice of
$f$-divergence. This provides new excess--housekeeping decompositions beyond
HS and MN.}
\par
This paper is organized as follows.
In Sec.~\ref{sec:preliminaries}, we first explain the setup \redtext{and some
basics of} stochastic thermodynamics.
We then introduce excess--housekeeping decompositions of EPR along with a
brief review of known results.
We also introduce the $f$-divergence required for our decomposition.
Section~\ref{sec:main_results} presents our main results.
We first describe the construction of the decomposition and its orthogonality,
and then show that it recovers the HS and MN decompositions for appropriate
choices and limits.
We next discuss the two physical meanings of the excess EPR and construct
physically meaningful decompositions \redtext{other than} the HS and MN
decompositions.
We close the section with two remarks: the condition on the $f$-divergence for
the decomposition to exist for any velocity and \redtext{state}, which
clarifies why the HS and MN decompositions are special, and an interpretation
of the decomposition as a weighted Helmholtz decomposition.
Section~\ref{sec:examples} numerically \redtext{demonstrates} our results
using diffusion on a one-dimensional ring with periodic boundary conditions.
Finally, Sec.~\ref{sec:discussion} summarizes the paper and discusses future
perspectives.

\section{Preliminaries}\label{sec:preliminaries}
\subsection{Stochastic thermodynamics}\label{sec:setup}
We consider an overdamped Langevin system at position $\bm x\in\Omega$,
\redtext{where $\Omega$ is a $d$-dimensional region in $\mathbb{R}^d$.
On the boundary, we impose no-flux condition or periodic boundary conditions.}
The system is in contact with a heat bath and a work source. 
The heat bath is in equilibrium at temperature $T$. 
With particle mobility $\mu$ and Boltzmann constant $\kB$, the diffusion
coefficient is given by $D=\mu\kB T$. 
The work source induces a conservative force $-\bm\nabla U_t(\bm{x})$
associated with a time-dependent potential $U_t(\bm{x})$, and a
nonconservative force $\bm{f}_t^{\rm nc}(\bm{x})$. 
The particle is generally out of equilibrium. We denote the probability
density to find the particle at position $\bm x$ at time $t$ by
$\rho_t(\bm{x})$, which obeys the Fokker--Planck equation
\begin{align}
    \partial_t\rho_t(\bm{x})&=-\bm\nabla\cdot\left[\bm
v_t(\bm{x})\rho_t(\bm{x})\right],\label{eq:FP}\\
    \bm v_t(\bm{x})&=-\mu[\bm\nabla U_t(\bm{x})-\bm f_t^{\rm
nc}(\bm{x})]-D\bm\nabla\ln\rho_t(\bm{x}).\label{eq:velocity}
\end{align}
Throughout, we impose periodic boundary conditions or the condition that the
probability density and the probability current vanish at infinity.\par
We now consider the steady state of Eq.~(\ref{eq:FP}) with the force fixed
at time $t$. 
In the absence of nonconservative forces, i.e., $\bm f_t^{\rm nc}(\bm{x})=0$,
the instantaneous \redtext{equilibrium state}
$\rho_t^{\rm eq}(\bm{x})\propto\exp[-\beta U_t(\bm{x})]$, with
$\beta=(\kB T)^{-1}$, is the steady state, for which the probability
current $\bm J_t(\bm{x})=\bm v_t(\bm{x})\rho_t(\bm{x})$ vanishes. On the other
hand, in the presence of nonconservative forces, the steady
solution of Eq.~(\ref{eq:FP}) is the nonequilibrium steady state
$\rho_t^{\rm st}(\bm{x})$ characterized by
\begin{align}
  0&=-\bm\nabla\cdot\left[\bm v_t^{\rm st}(\bm{x})\rho_t^{\rm
st}(\bm{x})\right],\label{eq:FP_st}\\
    \bm v_t^{\rm st}(\bm{x})&=-\mu[\bm\nabla U_t(\bm{x})-\bm f_t^{\rm
nc}(\bm{x})]-D\bm\nabla\ln\rho_t^{\rm st}(\bm{x}),\label{eq:velocity_st}
\end{align}
which generally carries a nonzero nonequilibrium steady current
$\bm J_t^{\rm st}(\bm{x})=\bm v_t^{\rm st}(\bm{x})\rho_t^{\rm st}(\bm{x})$.\par
We next introduce the entropy balance for this process. The entropy of the
system is given by $S_t\coloneqq-\int_\Omega\rho_t(\bm x)\ln\rho_t(\bm
x)\dd\bm x$. The rate of change in the entropy of the heat bath is given by
$-\beta\dot Q_t$ with the rate of heat absorbed by the system $\dot
Q_t\coloneqq\int_\Omega\bm v_t(\bm{x})\rho_t(\bm{x})\cdot[\bm\nabla U_t(\bm
x)-\bm f_t^{\rm nc}(\bm x)]\dd\bm x$. The EPR is then defined as the rate of
change in the entropy of the total system consisting of the particle and the
heat bath,
\begin{align}
    \dot\Sigma_t\coloneqq\dot S_t-\beta\dot Q_t,\label{eq:EP}
\end{align}
which quantifies the irreversibility of the process.\par
To clarify the geometric meaning of EPR, we define the inner
product of vector fields $\bm u,\ \bm w$ with respect to a general density
$\rho$ as $\langle\bm u,\bm w\rangle_{\rho}\coloneqq D^{-1}\int_\Omega\bm
u(\bm x)\cdot\bm w(\bm x)\rho(\bm x)\dd\bm x$, which we call the
$\rho$--weighted inner product. We also define the corresponding norm by
$\|\bm u\|_\rho^2\coloneqq \langle\bm u,\bm u\rangle_{\rho}$. Substituting
Eq.~(\ref{eq:FP}) into Eq.~(\ref{eq:EP}) and integrating by parts with the
boundary conditions introduced above, we can express the EPR as
\begin{align}
    \dot\Sigma_t=\|\bm v_t\|_{\rho_t}^2.\label{eq:EP_norm}
\end{align}
Thus, the EPR can be viewed as the squared norm of the
mean local velocity $\bm v_t(\bm x)$.
Note that the second law of thermodynamics, $\dot\Sigma_t\geq0$, follows
directly from Eq.~(\ref{eq:EP_norm}).
\subsection{Geometric decomposition of EPR}
Equation~(\ref{eq:EP_norm}) allows us to decompose EPR geometrically. The
general idea is to use an orthogonal decomposition of the mean local velocity
\begin{align}
    &\bm v_t(\bm x)=\redtext{\bm v_t^{\rm ex}}(\bm x)+\redtext{\bm v_t^{\rm
hk}}(\bm x),\label{eq:velocity_orthogonal}\\
    &\left\langle\redtext{\bm v_t^{\rm ex}},\redtext{\bm v_t^{\rm
hk}}\right\rangle_{\rho_t}=0.\label{eq:velocity_orthogonal_2}
\end{align}
\redtext{
We call the terms $\bm v_t^{\rm ex}$ and $\bm v_t^{\rm hk}$ the excess
velocity and the housekeeping velocity, respectively, and require $\bm
v_t^{\rm hk}=\bm 0$ whenever $\bm f_t^{\rm nc}=\bm
0$~\cite{DechantSasaIto2022PRR}. 
The housekeeping velocity can thus be regarded as the velocity required to
maintain the nonequilibrium nature arising from the nonconservative force, and
the excess velocity as the additional velocity required beyond it.
}\par
The orthogonal decomposition of the velocity \eqref{eq:velocity_orthogonal}
yields the Pythagorean decomposition of the EPR
\begin{align}
    \dot\Sigma_t=\left\|\redtext{\bm v_t^{\rm
ex}}\right\|_{\rho_t}^2+\left\|\redtext{\bm v_t^{\rm
hk}}\right\|_{\rho_t}^2.\label{eq:general_decomposition}
\end{align}
\redtext{The first and second terms are called the excess EPR and housekeeping
EPR, respectively, and represent the entropic costs associated with the excess
and housekeeping velocities, respectively.
When the orthogonality~\eqref{eq:velocity_orthogonal_2} and the property that
$\bm v_t^{\rm hk}=\bm 0$ for $\bm f_t^{\rm nc}=\bm 0$ hold, we call the
decomposition of the velocity~\eqref{eq:velocity_orthogonal_2} and the
decomposition of the EPR~\eqref{eq:general_decomposition} excess--housekeeping
decompositions. These requirements do not uniquely determine the
decomposition, and infinitely many choices are possible.}
In what follows, we discuss two established approaches.\par
\subsubsection{\redtext{Hatano--Sasa} decomposition}
\begin{figure}[t]
 \centering
 \includegraphics[width=180pt]
 {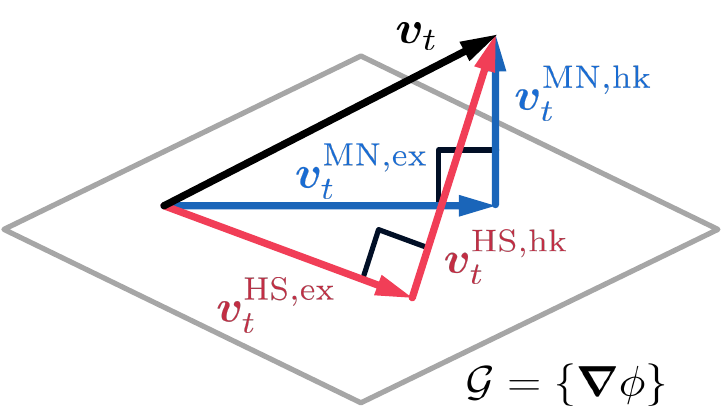}
 \caption{Schematic illustration of the excess--housekeeping decomposition. In
both the HS and MN decompositions, the excess velocity lies in the space of
gradient fields $\mathcal{G}$. In particular, the MN decomposition gives an
orthogonal projection. In the HS decomposition, the excess velocity is denoted
by $\bm v_t^{\rm HS,ex}$ and the housekeeping velocity by $\bm v_t^{\rm
HS,hk}$; in the MN decomposition, the excess velocity is denoted by $\bm
v_t^{\rm MN,ex}$ and the housekeeping velocity by $\bm v_t^{\rm MN,hk}$.
 }
 \label{fig:projection}
\end{figure}

The first approach is based on the nonequilibrium steady state and called the
Hatano--Sasa (HS) decomposition~\cite{HatanoSasa2001,Ge2009,VanDenBroeckEsposi%
to2010FP}.
Computing the $\rho_t$--weighted inner product between the steady velocity
$\bm v_t^{\rm st}$ and $\bm v_t-\bm v_t^{\rm st}$ using the definition of the
steady state~(\ref{eq:FP_st}) yields $\left\langle\bm v_t-\bm v_t^{\rm st},\bm
v_t^{\rm st}\right\rangle_{\rho_t}=0$. Thus, $\bm v_t=(\bm v_t-\bm v_t^{\rm
st})+\bm v_t^{\rm st}$ is an orthogonal decomposition. \redtext{On the basis
of} this orthogonality, we can construct the HS decomposition of the EPR:
\begin{align}
    &\dot\Sigma_t=\dot\Sigma_t^{\rm HS,ex}+\dot\Sigma_t^{\rm HS,hk},\\
    &\dot\Sigma_t^{\rm HS,ex}\coloneqq\left\|\bm v_t-\bm v_t^{\rm
st}\right\|_{\rho_t}^2,\\
    &\dot\Sigma_t^{\rm HS,hk}\coloneqq\left\|\bm v_t^{\rm
st}\right\|_{\rho_t}^2.
\end{align}
 Note that the excess velocity can be written as
$\bm v_t^{\rm HS, ex}\coloneqq \bm v_t-\bm v_t^{\rm
st}=-D\bm\nabla\ln(\rho_t/\rho_t^{\rm st})$.
Hence, the excess velocity belongs to the space of all
gradient fields $\mathcal{G}
  \coloneqq
  \left\{
    \bm\nabla\phi
    \,\middle|\,
    \phi:\Omega\to\mathbb{R}
    \text{ is a smooth scalar function}\right\}$
(Fig.~\ref{fig:projection}).\par
The excess EPR $\dot\Sigma_t^{\rm HS,ex}$ is interpreted as the cost
associated with relaxation toward the instantaneous steady state. Indeed, the
excess EPR coincides with the rate of decrease in the KL divergence with
respect to the steady \redtext{state}, $\mathcal{D}_{\rm
KL}(\rho_t\|\rho_t^{\rm st})\coloneqq\int_\Omega\rho_t(\bm
x)\ln\frac{\rho_t(\bm x)}{\rho_t^{\rm st}(\bm x)}\dd\bm
x$~\cite{VanDenBroeckEsposito2010FP}:
\begin{align}
    \dot\Sigma_t^{\rm HS,ex}=-\left.\frac{\partial}{\partial
s}\mathcal{D}_{\rm KL}(\rho_s\|\rho_t^{\rm
st})\right|_{s=t}.\label{eq:HS_excess_KL}
\end{align}
The housekeeping EPR $\dot\Sigma_t^{\rm HS,hk}$ is interpreted as the cost of
maintaining the instantaneous steady current.
When $\bm f_t^{\rm nc}(\bm x)=0$, we have $\bm v_t^{\rm st}(\bm x)=0$ and thus
$\dot\Sigma_t^{\rm HS,hk}=0$.
\subsubsection{\redtext{Maes--Neto{\v{c}}n{\'y}} decomposition}
The second approach is a formulation based on optimal transport theory and is
called the Maes--Neto{\v{c}}n{\'y} (MN) decomposition.
Given $\rho_t$ and $\bm v_t$, we consider the mean local velocity $\bm u$ that
minimizes the EPR among those yielding the same time
evolution \redtext{as $\partial_t\rho_t=-\bm\nabla\cdot(\rho_t\bm v_t)$}:
\begin{align}
    \bm v_t^*\coloneqq\underset{\substack{\bm u\\
  \bm\nabla\cdot(\rho_t\bm u)=\bm\nabla\cdot(\rho_t\bm v_t)}}
  {\operatorname{arg\,min}}\,\|\bm u\|_{\rho_t}^{2}.\label{eq:MN_OT}
\end{align}
This problem is known to be equivalent to a short-time optimal transport
problem~\redtext{\cite{DechantSasaIto2022PRR}}, and its unique optimal
solution can be written as
$\bm v_t^*(\bm x)=\bm\nabla\phi_t^{\rm MN}$ in terms of the potential
$\phi_t^{\rm MN}$ satisfying
\begin{align}
    &\bm\nabla\cdot[\bm v_t(\bm x)\rho_t(\bm
x)]=\bm\nabla\cdot\left[\bm\nabla\phi_t^{\rm MN}(\bm x)\rho_t(\bm
x)\right],\notag\\
    &\int_\Omega\rho_t(\bm x)\phi_t^{\rm MN}(\bm x)\dd\bm
x=0.\label{eq:orth_MN}
\end{align}
Since
$\left\langle\bm v_t-\bm\nabla\phi_t^{\rm MN},\bm\nabla\phi_t^{\rm
MN}\right\rangle_{\rho_t}=0$
holds, we have an orthogonal decomposition of the velocity 
$\bm v_t=\left(\bm v_t-\bm \nabla\phi_t^{\rm MN}\right)+\bm\nabla\phi_t^{\rm
MN}$.
\redtext{On the basis of this orthogonality,} we can construct the MN
decomposition of
the EPR,
\begin{align}
    &\dot\Sigma_t=\dot\Sigma_t^{\rm MN,ex}+\dot\Sigma_t^{\rm MN,hk},\\
    &\dot\Sigma_t^{\rm MN,ex}\coloneqq\left\|\bm\nabla\phi_t^{\rm
MN}\right\|_{\rho_t}^2,\\
    &\dot\Sigma_t^{\rm MN,hk}\coloneqq\left\|\bm v_t-\bm\nabla\phi_t^{\rm
MN}\right\|_{\rho_t}^2.
\end{align}
Here, $\dot\Sigma_t^{\rm MN,ex}$ represents the minimal cost of changing the
\redtext{state}, whereas $\dot\Sigma_t^{\rm MN,hk}$ is interpreted as the cost
of
the velocity part that does not change the \redtext{state}.\par
\redtext{
We now give a geometric interpretation of the MN decomposition.
Using Eq.~(\ref{eq:orth_MN}), we have
$\left\langle\bm v_t-\bm\nabla\phi_t^{\rm MN},
\bm\nabla\phi\right\rangle_{\rho_t}=0$
for any scalar function $\phi$.
Since the MN excess velocity $\bm\nabla\phi_t^{\rm MN}$
belongs to $\mathcal G$, it is the orthogonal projection
of $\bm v_t$ onto $\mathcal G$.
}\par
As we have seen, the HS and MN excess EPRs admit representations with
different physical interpretations, given in Eqs.~\eqref{eq:HS_excess_KL}
and~\eqref{eq:MN_OT}, respectively.
However, there has been no framework that unifies the two decompositions while
explicitly treating those properties.
Using $f$-divergence as a mathematical tool, we unify them under a common
thermodynamic interpretation, in which the excess part is regarded as the EPR
associated with relaxation toward a target distribution, and the housekeeping
part as the cost of maintaining the flow that preserves the same relaxation
target.
\subsection{$f$-divergence}
Before presenting our main result, we review \redtext{some} basic properties
of the $f$-divergence.
For probability densities $\rho,\ \gamma$ and a smooth convex
function $f:(0,\infty)\to\mathbb R$ satisfying $f(1)=f'(1)=0$, we define the
$f$-divergence as
\begin{align}
    \mathcal{D}_f(\rho\|\gamma)\coloneqq\int_\Omega f\left(\frac{\rho(\bm
x)}{\gamma(\bm x)}\right)\gamma(\bm x)\dd\bm x.
\end{align}
This quantity measures the difference between the two probability densities
$\rho$ and $\gamma$.
The function $f$ is called the generating function.
The condition $f'(1)=0$ is imposed to remove the ambiguity of generating
functions that yield the same divergence.\par
\redtext{Applications of general $f$-divergences include estimates of rates of
convergence to steady states for Fokker--Planck
equations~\cite{ArnoldEtAl2001} and lower bounds on the Bayes risk in
statistical estimation~\cite{ChenGuntuboyinaZhang2016}.
Specific $f$-divergences have also found applications in statistical physics.
For example, $f(r)=r\ln r-r+1$ yields the Kullback--Leibler divergence, which
relates the free energy of a nonequilibrium state to that of the corresponding
equilibrium state~\cite{EspositoVanDenBroeck2011Landauer}.
The choice $f(r)=-\ln r+r-1$ yields the reverse Kullback--Leibler divergence,
which is minimized to train Boltzmann generators for equilibrium
sampling~\cite{KohlerEtAl2023RigidBodyFlows}.
The choice $f(r)=(\sqrt{r}-1)^2$ yields the squared Hellinger divergence,
which has been used to derive thermodynamic uncertainty
relations~\cite{Hasegawa2023BulkBoundary}.}

\begin{figure*}[t]
 \centering
 \includegraphics[width=480pt]{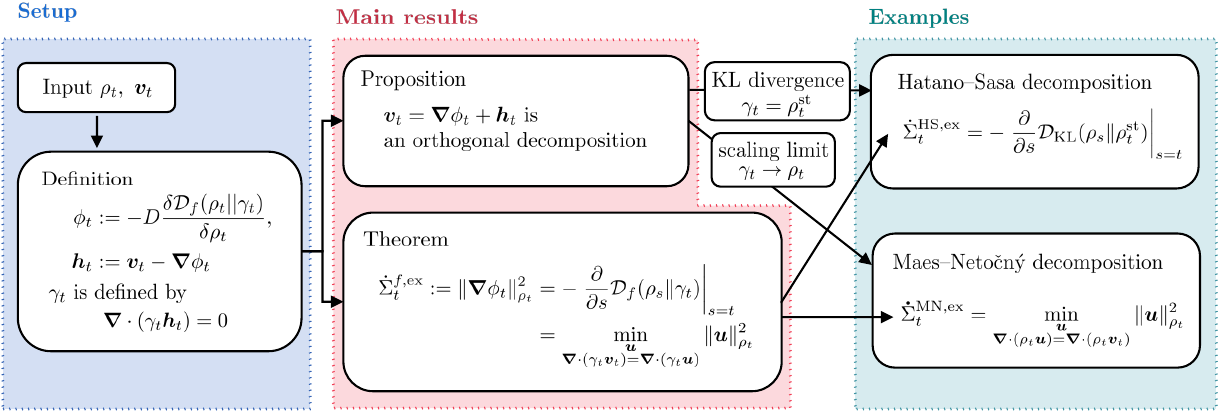}
 \caption{
 Logical structure of the decomposition proposed in this work. For given
$\rho_t,\ \bm v_t$, we first define $\phi_t, \bm h_t,\ \gamma_t$ (blue box).
Our main \redtext{results} then establish the orthogonality of the
decomposition and two representations of the excess EPR (red box). Under
appropriate settings, these recover the HS and MN decompositions (green box).
 }
 \label{fig:logic}
\end{figure*}

Consider the \(f\)-divergence \(\mathcal{D}_f(\rho_t\|\gamma)\) between a
fixed $\gamma$ and a density $\rho_t$ evolving according to
$\partial_t\rho_t=-\bm\nabla\cdot(\bm u_t\rho_t)$.
The instantaneous velocity $\bm u_t$ that decreases
\(\mathcal{D}_f(\rho_t\|\gamma)\) most steeply per unit norm is called the
gradient flow of \(\mathcal{D}_f(\rho_t\|\gamma)\).
This concept plays an important role in, e.g., the variational formulation of
the Fokker--Planck equation~\cite{JordanKinderlehrerOtto1998} and nonlinear
diffusion driven by general entropy
functionals~\cite{Otto2001,CarrilloMcCannVillani2003}.
When the geometry of probability densities is specified by the
$\rho_t$--weighted inner product $\langle\cdot,\cdot\rangle_{\rho_t}$, the
first variation \(\delta
\mathcal{D}_f(\rho_t\|\gamma)/\delta\rho_t=f'(\rho_t/\gamma)\) for fixed
\(\gamma\) yields the gradient flow~\cite{AmbrosioGigliSavare2005}
\begin{align}
  \bm u_t
  &=
  -D\bm\nabla f'\!\left(\frac{\rho_t}{\gamma}\right).
  \label{eq:f-divergence-gradient-potential}
\end{align}
\parhead{Main result}\label{sec:main_results}\redtext{With} these
preliminaries, we now present our main result. Using an $f$-divergence, we
construct the velocity decomposition $\bm v_t(\bm x)=\bm\nabla\phi_t(\bm
x)+\bm h_t(\bm x)$, which is orthogonal and yields an excess--housekeeping
decomposition of the EPR. The excess part $\bm\nabla\phi_t(\bm x)$ describes
relaxation toward an instantaneous relaxation target measured by a specified
$f$-divergence, while the housekeeping part $\bm h_t(\bm x)$ preserves the
relaxation target. %
Figure~\ref{fig:logic} summarizes the logical structure of the main results in
this section.
\subsection{Construction and orthogonality}
We begin by describing how the decomposition is constructed.
The decomposition is determined by a given probability density $\rho_t(\bm
x)$, velocity $\bm v_t(\bm x)$, and prescribed $f$-divergence.
We first define the potential function $\phi_t$, the velocity $\bm h_t$, and
the probability density $\gamma_t$ that are required to define the
decomposition.
\begin{defi*}
    For a probability density $\rho_t(\bm x)$, velocity $\bm v_t(\bm x)$, and
$f$-divergence $\mathcal{D}_f(\cdot\|\cdot)$, we define a potential function
$\phi_t$, velocity $\bm h_t$, and probability density $\gamma_t$ by the
following three equations:
    \begin{align}
    &\phi_t(\bm x)
    \coloneqq-D\frac{\delta \mathcal{D}_f(\rho_t\|\gamma_t)}{\delta\rho_t(\bm
x)}
    =-D f'\left(\frac{\rho_t(\bm x)}{\gamma_t(\bm x)}\right),
    \label{eq:condition_a}\\
    &\bm h_t(\bm x)\coloneqq\bm v_t(\bm x)-\bm\nabla\phi_t(\bm
x),\label{eq:housekeeping_def}\\
    &\bm\nabla\cdot[\bm h_t(\bm x)\gamma_t(\bm x)]
    =0.
    \label{eq:condition_b}
\end{align}
For no-flux boundary conditions, we additionally impose $(\gamma_t\bm
h_t)\cdot\bm n=0$ on $\partial\Omega$, where $\bm n$ is the outward unit
normal.
\end{defi*}
We explain the meaning of each equation. First, Eq.~\eqref{eq:condition_a}
expresses that the potential $\phi_t$ describes the relaxation of the
$f$-divergence $\mathcal{D}_f(\rho_t\|\gamma_t)$ between \redtext{the current
state $\rho_t$ and the relaxation target $\gamma_t$, which is} unknown at this
stage.
This means that the gradient $\bm\nabla\phi_t$ coincides with the
gradient-flow velocity of $\mathcal{D}_f(\rho_t\|\gamma_t)$. In our
decomposition, $\bm\nabla\phi_t$ corresponds to the excess velocity.
\redtext{Equations~\eqref{eq:housekeeping_def} and~\eqref{eq:condition_b}
state that the residual part $\bm h_t$ leaves the target state $\gamma_t$
unchanged and thus serves as the housekeeping velocity.
The target state $\gamma_t$ is then determined by solving the system of
equations~\eqref{eq:condition_a}--\eqref{eq:condition_b} for the given
$\rho_t$, $\bm v_t$, and $f$-divergence.}\par
For the quantities defined above, the following \redtext{proposition} holds.
This \redtext{proposition} provides the basis for the decomposition presented
in this work.
\begin{prop}\label{prop:orthogonality}
   If there exists a relaxation target $\gamma_t$ satisfying
Eqs.~\eqref{eq:condition_a},~\eqref{eq:housekeeping_def},
and~\eqref{eq:condition_b}, then the corresponding velocity decomposition
   \begin{align}
       \bm v_t(\bm x)=\bm\nabla\phi_t(\bm x)+\bm h_t(\bm x)
   \end{align}
   is an orthogonal decomposition with respect to the $\rho_t$--weighted inner
product.
\end{prop}
\begin{proof}[\redtext{Proof}]
Using the density ratio $r_t(\bm x)\coloneqq \rho_t(\bm x)/\gamma_t(\bm x)$,
we introduce the auxiliary potential $\psi_t(\bm x)\coloneqq D[f(r_t(\bm
x))-r_t(\bm x)f'(r_t(\bm x))]$. Equation~(\ref{eq:condition_a}) then gives 
\begin{align}
    \rho_t(\bm x)\bm\nabla\phi_t(\bm x)=\gamma_t(\bm x)\bm\nabla\psi_t(\bm
x).\label{eq:rho_gamma}
\end{align}
Additionally, Eq.~\eqref{eq:condition_b} implies that $\bm h_t$ is orthogonal
to the gradient of any potential function with respect to the
$\gamma_t$--weighted inner product. Combining these results, we obtain
\begin{align}
    \langle\bm\nabla\phi_t,\bm h_t\rangle_{\rho_t}
    &=\langle\bm\nabla\psi_t,\bm h_t\rangle_{\gamma_t}=0,
\end{align}
which proves the orthogonality.
\end{proof}
\redtext{In the absence of nonconservative forces $f^{\rm nc}_t(x) = 0$, we
can show $h_t = 0$, as expected for the housekeeping part.}
This follows because $\bm h_t$ is orthogonal to the gradient of any potential
function with respect to the $\gamma_t$--weighted inner product and $\bm h_t$
in this case itself takes the gradient form $\bm h_t(\bm x)=-\bm\nabla(\mu
U_t+D\ln\rho_t+\phi_t)$.
\par
From the orthogonality and the correspondence between $\bm h_t$ and
nonconservativeness, the decomposition $\bm v_t(\bm x)=\bm\nabla\phi_t(\bm
x)+\bm h_t(\bm x)$ can be regarded as an excess--housekeeping decomposition of
the velocity.
Accordingly, the EPR also admits the excess--housekeeping decomposition
\begin{align}
    &\dot\Sigma_t=\dot\Sigma_t^{f,\rm ex}+\dot\Sigma_t^{f,\rm
hk},\label{eq:f_decomposition}\\
    &\dot\Sigma_t^{f,\rm
ex}\coloneqq\left\|\bm\nabla\phi_t\right\|_{\rho_t}^2,\\
    &\dot\Sigma_t^{f,\rm hk}\coloneqq\left\|\bm h_t\right\|_{\rho_t}^2.
\end{align}
\subsection{Recovery of existing decompositions}
We next show that the HS and MN decompositions are recovered within our
construction by considering two particular situations.\par
\redtext{The HS decomposition is recovered by choosing} the generating
function that yields the KL divergence:
\begin{align}
 f_{\rm KL}(r)
 &\coloneqq
 r\ln r-r+1.
\label{eq:scaled_kl_generator}
\end{align}
Indeed, from Eq.~\eqref{eq:condition_a}, the excess potential in this case can
be written as \(\phi_t^{\rm KL}=-D\ln(\rho_t/\gamma_t^{\rm KL})\) with the
corresponding relaxation target $\gamma_t^{\rm KL}$, and the housekeeping
velocity \(\bm h_t^{\rm KL}\) is given by
\begin{align}
 \bm h_t^{\rm KL}
 &=
 -\mu
 \left(
 \bm\nabla U_t-\bm f_t^{\rm nc}
 \right)
 -
 D\bm\nabla\ln\gamma_t^{\rm KL}.
\label{eq:HS_housekeeping}
\end{align}
 Then, Eq.~\eqref{eq:condition_b} for this velocity is equivalent to the
stationarity condition, Eq.~(\ref{eq:FP_st}).
Therefore, $\gamma_t^{\rm KL}=\rho_t^{\rm st}$ is a solution, and $\bm
h_t^{\rm KL}=\bm v_t^{\rm st}$ recovers the HS decomposition.\par
The MN decomposition is recovered in the limit in which the relaxation target
$\gamma_t$
converges to the current \redtext{state} $\rho_t$.
\redtext{Such a limit is obtained by scaling} an arbitrary generating function
$f$ satisfying $f''(1)>0$ with a parameter \(\epsilon>0\) as
\begin{align}
 f_\epsilon(r)
 &\coloneqq
 \frac{f(r)}{\epsilon}.
\label{eq:mn_scaled_family}
\end{align}
We denote the $f$-divergence defined by this generating function by
$\mathcal{D}_{f_\epsilon}(\cdot\|\cdot)$, the corresponding relaxation target
by $\gamma_t^\epsilon$, and the decomposition by $\bm
v_t=\bm\nabla\phi_t^{\epsilon}+\bm h_t^{\epsilon}$. \redtext{Under some mild
assumptions detailed in Appendix~\ref{app:mn_limit}, $\gamma_t^\epsilon$
converges to $\rho_t$ as $\epsilon\to0$.
In addition, $\phi_t^\epsilon$ and $\bm h_t^\epsilon$ have corresponding
limits, denoted by $\phi_t^0$ and $\bm h_t^0$, respectively, where $\phi_t^0$
satisfies}
\begin{align}
    \bm\nabla\cdot[\bm v_t(\bm x)\rho_t(\bm
x)]&=\bm\nabla\cdot[\bm\nabla\phi_t^0(\bm x)\rho_t(\bm x)],\notag\\
    \int_\Omega\rho_t(\bm x)\phi_t^0(\bm x)\dd\bm x&=0.
\end{align}
Since this is equivalent to Eq.~\eqref{eq:orth_MN}, we have
$\phi_t^0=\phi_t^{\rm MN}$, and the MN decomposition is recovered in this
limit. The detailed derivation is given in Appendix~\ref{app:mn_limit}.

We also find that the MN \redtext{excess EPR} provides
an upper bound on \redtext{other excess EPRs} in our construction. By
Eq.~\eqref{eq:orth_MN},
\(\bm v_t-\bm\nabla\phi_t^{\rm MN}\) is orthogonal to any gradient field
with respect to the $\rho_t$--weighted inner product. Thus, for any generating
function $f$ and the corresponding excess potential $\phi_t$, we have
$\langle\bm\nabla\phi_t^{\rm MN},\bm\nabla\phi_t\rangle_{\rho_t}=\langle\bm
v_t,\bm\nabla\phi_t\rangle_{\rho_t}=\|\bm\nabla\phi_t\|_{\rho_t}^2$. This gives
\begin{align}
 \dot\Sigma_t^{\rm MN,ex}-\dot\Sigma_t^{f,\rm ex}
 &= \|\bm\nabla\phi_t^{\rm MN}-\bm\nabla\phi_t\|_{\rho_t}^2\geq0.
 \label{eq:mn_excess_upper_bound}
\end{align}
Thus, the MN decomposition gives the largest excess EPR within our
construction.
\subsection{Physical interpretation of the excess part}
We next explain \redtext{that the excess EPR in our construction unifies the
two physical interpretations in terms of relaxation and optimal transport, as
shown in the following theorem.}
\begin{thm}\label{thm:physical_interpretations}
    \redtext{In the decomposition~\eqref{eq:f_decomposition}, the excess EPR
has the following two representations:}
    \begin{align}
    \dot\Sigma_t^{f,\rm ex}&=-\left.\frac{\partial}{\partial
s}\mathcal{D}_f(\rho_s\|\gamma_t)\right|_{s=t}\label{eq:excess_differential}\\
    &=
  \min_{\substack{\bm u\\
  \bm\nabla\cdot(
  \gamma_t\bm v_t)=\bm\nabla\cdot(
  \gamma_t\bm u)}}
  \|\bm u\|_{\rho_t}^{2}\redtext{.}
  \label{eq:f_minimum_action}
\end{align}
Under no-flux boundary conditions, all minimizations over velocities are
restricted to fields satisfying $\bm u\cdot\bm n=0$ on $\partial\Omega$.
\end{thm}
These two expressions, Eqs.~\eqref{eq:excess_differential}
and~\eqref{eq:f_minimum_action}, provide the two representative aspects of our
decomposition. The first expression, Eq.~\eqref{eq:excess_differential},
states that the excess EPR equals \redtext{the $f$-divergence with respect to
the target $\gamma_t$. This generalizes} the physical interpretation of the
excess EPR in the HS decomposition given by Eq.~\eqref{eq:HS_excess_KL}. The
second expression, Eq.~\eqref{eq:f_minimum_action}, gives a variational
representation of the excess EPR \redtext{based on optimal transport},
generalizing the \redtext{physical interpretation of} the MN decomposition in
Eq.~\eqref{eq:MN_OT}. {Thus, our decomposition unifies the relaxation
interpretation of the HS decomposition and the optimal transport
interpretation of the MN decomposition under the common structure that the
housekeeping part preserves the same relaxation target $\gamma_t$
(Fig.~\ref{fig:decomp}).
\begin{figure}[t]
 \centering
 \includegraphics[width=200pt]
 {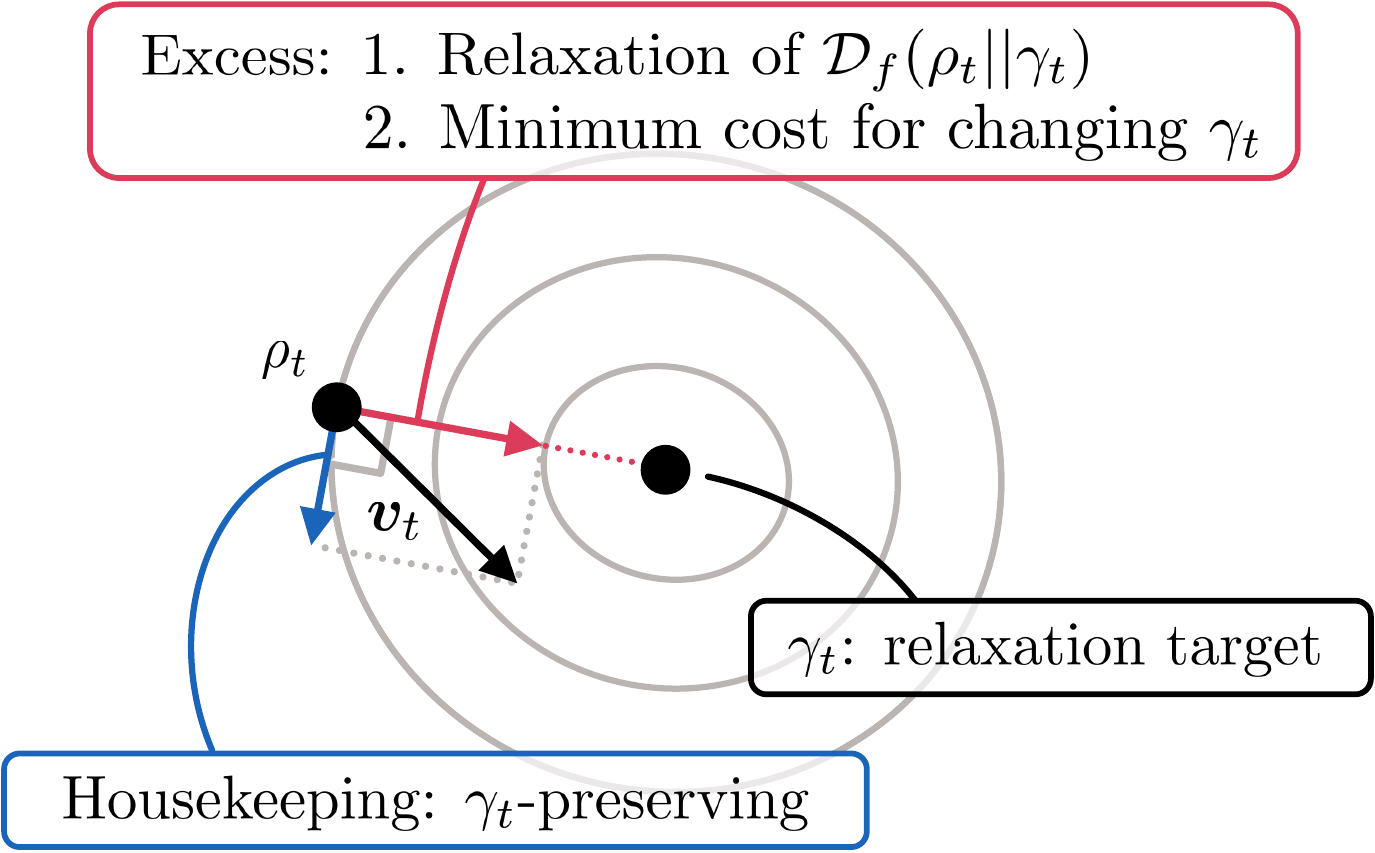}
 \caption{Schematic illustration of our decomposition and its physical
interpretations. For the \redtext{instantaneous state} $\rho_t$, the velocity
$\bm v_t$ driving its evolution is orthogonally decomposed into excess and
housekeeping parts. The excess part describes both relaxation measured by
$\mathcal{D}_f(\rho_t\|\gamma_t)$ and the minimum cost of evolving the
relaxation target $\gamma_t$, whereas the housekeeping part preserves
$\gamma_t$. The gray curves are contours of $\mathcal{D}_f(\cdot\|\gamma_t)$.
 }
 \label{fig:decomp}
\end{figure}
In what follows, we \redtext{derive} Eqs.~\eqref{eq:excess_differential}
and~\eqref{eq:f_minimum_action} separately.}\par
\redtext{We first derive Eq.~\eqref{eq:excess_differential}.} Using the
Fokker--Planck equation~(\ref{eq:FP}), we calculate the rate of decrease of
the $f$-divergence as
\begin{align}
    -\left.\frac{\partial}{\partial
s}\mathcal{D}_f(\rho_s\|\gamma_t)\right|_{s=t}
    &=-\frac{1}{D}\int_\Omega\phi_t(\bm x)\bm\nabla\cdot[\bm v_t(\bm
x)\rho_t(\bm x)]\dd\bm x\\
    &=\langle\bm\nabla\phi_t,\bm v_t\rangle_{\rho_t}\\
    &=\left\|\bm\nabla\phi_t\right\|_{\rho_t}^2\\
    &=\dot\Sigma_t^{f,\rm ex}.
\end{align}
\par
Here, we consider the case $f=f_{\rm KL}$. As shown in the previous section,
the decomposition then reduces to the HS decomposition, and we obtain
$\gamma_t=\rho^{\rm st}_t$.
Equation~\eqref{eq:excess_differential} then coincides with
Eq.~(\ref{eq:HS_excess_KL}), a well-known property of the HS decomposition.
Moreover, since the MN decomposition is obtained in the limit $\epsilon\to+0$
for $f_\epsilon$ defined in Eq.~\eqref{eq:mn_scaled_family}, we obtain the
relaxational representation of the MN excess EPR,
\begin{align}
    \dot{\Sigma}_t^{\rm
MN,ex}=-\lim_{\epsilon\to+0}\left.\frac{\partial}{\partial
s}\mathcal{D}_{f_\epsilon}(\rho_s\|\gamma_t^\epsilon)\right|_{s=t}.
\end{align}
\par
Note that \redtext{Ref.~\cite{DechantSasaIto2022PRE}} showed that the MN
excess EPR can be expressed as the instantaneous rate of decrease of the KL
divergence relative to a certain reference density.
However, the reference density used there is in general not preserved by the
MN housekeeping part.
In contrast, our framework expresses the MN excess EPR as the
\redtext{(limiting)} relaxation rate of an $f$-divergence relative to the
relaxation target $\gamma_t$, which is preserved by the housekeeping part $\bm
h_t$.
\par
We next \redtext{derive} the second expression~\eqref{eq:f_minimum_action}.
For an arbitrary velocity field $\bm u$, let
$\bm w\coloneqq\bm u-\bm\nabla\phi_t$. If $\bm u$ satisfies
$\bm\nabla\cdot[\gamma_t(\bm v_t-\bm u)]=0$, substituting
condition~\eqref{eq:condition_b} gives $\bm\nabla\cdot(\gamma_t\bm w)=0$.
Hence, $\bm w$ is orthogonal to any gradient field with respect to the
$\gamma_t$--weighted inner product.
Using Eq.~\eqref{eq:rho_gamma}, we obtain the orthogonality relation
$\langle\bm\nabla\phi_t,\bm w\rangle_{\rho_t}=\langle\bm\nabla\psi_t,\bm
w\rangle_{\gamma_t}=0$. Therefore,
\begin{align}
  \|\bm u\|_{\rho_t}^{2}
  &=\|\bm\nabla\phi_t\|_{\rho_t}^{2}+\|\bm
w\|_{\rho_t}^{2}\geq\dot{\Sigma}_t^{f,\mathrm{ex}}.
\end{align}
The equality holds for $\bm u=\bm\nabla\phi_t$. By Eq.~\eqref{eq:condition_b},
this velocity satisfies the constraint $\bm\nabla\cdot[\gamma_t(\bm v_t-\bm
u)]=0$, which proves the variational expression~\eqref{eq:f_minimum_action}
for the excess EPR. Thus, the excess velocity in our decomposition has the
minimum norm among the velocities that induce the same instantaneous time
evolution as the original velocity $\bm v_t$ when acting on $\gamma_t$.\par
This variational representation of the excess EPR in terms of optimal
transport recovers the MN optimal transport representation~\eqref{eq:MN_OT} in
the limit $\gamma_t\to\rho_t$. For $\gamma_t=\rho_t^{\rm st}$,
\redtext{corresponding to} the HS decomposition, we obtain the variational
representation for the HS excess EPR,
\begin{align}
    \dot{\Sigma}^{\rm HS,ex}_t=\min_{\substack{\bm u\\
  \bm\nabla\cdot(
  \rho_t^{\rm st}\bm v_t)=\bm\nabla\cdot(
  \rho_t^{\rm st}\bm u)}}
  \|\bm u\|_{\rho_t}^{2},
\end{align}
which is consistent with the projection representation derived in
Ref.~\cite{DechantSasaIto2022PRE}.
\subsection{Other examples of our decomposition}
\label{subsec:another_decompositions}
In this subsection, we construct examples of decompositions in our class other
than the HS and MN decompositions.
When a \redtext{state} is prepared by applying external forces, it is
important to evaluate not only the relaxation toward the actual steady
\redtext{state} but also the dissipation associated with the approach to a
desired \redtext{state}~\cite{AurellEtAl2012}.
From this viewpoint, we assume the presence of uncontrollable nonconservative
forces or uncontrollable temperature changes. 
We then choose $f$ such that $\gamma_t$ coincides with the desired steady
\redtext{state}, i.e., that without uncontrollable nonconservative forces or
temperature changes.
In this way, we construct a method to decompose the EPR into a part associated
with the change toward the desired \redtext{state} and a part that preserves
\redtext{that state}.
\subsubsection{Equilibrium-based decomposition}
\label{subsubsec:another_equilibrium}
First, we consider the situation in which an uncontrollable nonconservative
force $\bm f_t^{\rm nc}$ acts in addition to the controllable conservative
force $-\bm\nabla U_t$.
In this case, the steady \redtext{state} assumed by the controller is the
instantaneous equilibrium \redtext{state} $q_t\coloneqq Z_t^{-1}e^{-\beta
U_t}$, with $Z_t\coloneqq\int_\Omega e^{-\beta U_t}\dd\bm x$.
\par
We now explain how to construct the decomposition such that this
\redtext{state} coincides with the relaxation target $\gamma_t$.
Let $\bm\nabla\chi_t$ denote the gradient part of the velocity induced by the
nonconservative force, where $\chi_t$ satisfies $\int_\Omega q_t\chi_t\dd\bm
x=0$ and
\begin{align}
 \bm\nabla\cdot(q_t\bm\nabla\chi_t)
 &=\bm\nabla\cdot(q_t\mu\bm f_t^{\rm nc}).
 \label{eq:another_eq_poisson}
\end{align}
Requiring $\gamma_t=q_t$ fixes the excess velocity to be
$\bm\nabla\chi_t-D\bm\nabla\ln r_t$ with $r_t\coloneqq\rho_t/q_t$.
Therefore, if there exists a smooth function $K_{\rm eq}$ satisfying $K_{\rm
eq}(1)=0$, $1-rK_{\rm eq}'(r)\geq0$, and
\begin{align}
 \chi_t(\bm x)&=D K_{\rm eq}(r_t(\bm x))+C_t,
 \label{eq:another_eq_generator}
\end{align}
where $C_t$ is a spatially uniform constant, then the decomposition with
$\gamma_t=q_t$ is constructed by choosing the generating function $f=f_{\rm
eq}$ given by
\begin{align}
 f_{\rm eq}(r)&=f_{\rm KL}(r)-\int_1^r K_{\rm eq}(s)\dd s.
 \label{eq:another_eq_f_definition}
\end{align}

In this case, $f(1)=f'(1)=0$ and $f''(r)=r^{-1}-K_{\rm eq}'(r)\geq0$, and we
obtain $\bm\nabla\phi_t=\bm\nabla\chi_t-D\bm\nabla\ln r_t$ and $\bm h_t=\mu\bm
f_t^{\rm nc}-\bm\nabla\chi_t$.
Since $\bm\nabla\cdot(q_t\bm h_t)=0$ from Eq.~\eqref{eq:another_eq_poisson},
Eqs.~\eqref{eq:condition_a} and~\eqref{eq:condition_b} are satisfied.
For example, $K_{\rm eq}(r)=(1-c)\ln r$ with $c>0$ yields $f_{\rm eq}=cf_{\rm
KL}$.
Note that, in this construction, an appropriate $f$ does not necessarily exist
for arbitrary $\rho_t$ and $\bm v_t$.
The condition for its existence is discussed in
Appendix~\ref{app:equilibrium_reference}.
\par
This construction can be interpreted as evaluating the change toward a desired
equilibrium \redtext{state} in the presence of an uncontrollable external
flow. Under the external flow due to the nonconservative force $\bm f_t^{\rm
nc}$, the potential $U_t$ and temperature $T$ define the desired equilibrium
\redtext{state} $q_t=Z_t^{-1}e^{-\beta U_t}$. When the condition in
Eq.~\eqref{eq:another_eq_generator} is satisfied, we denote the corresponding
excess EPR and housekeeping EPR by $\dot\Sigma_t^{\rm eq,
ex}$,\ $\dot\Sigma_t^{\rm eq, hk}$, respectively. Then $\dot\Sigma_t^{\rm
eq,ex}$ represents the dissipation associated with the approach to the desired
equilibrium \redtext{state},
\begin{align}
 \dot\Sigma_t^{\rm eq,ex}
 &=-\left.
 \frac{\partial}{\partial s}
 \mathcal D_{f_{\rm eq}}(\rho_s\|q_t)
 \right|_{s=t}
 \geq0,
\end{align}
while $\dot\Sigma_t^{\rm eq,hk}$ represents the dissipation associated with
the flow that preserves that equilibrium \redtext{state}. Whereas the HS
decomposition uses the steady \redtext{state} of the system including the
external flow as its relaxation target, this decomposition evaluates the
approach to a prescribed equilibrium \redtext{state} along the actual process
in the presence of the external flow while retaining orthogonality. We refer
to this as the equilibrium-based decomposition.

\subsubsection{Temperature-based decomposition}
\label{subsubsec:another_temperature}
{We next consider a mismatch between the actual temperature $T$ and the
intended temperature $T'=\theta T$, with $\theta>0$ and $\theta\neq1$. In this
case, the instantaneous steady \redtext{state $q_t^\theta$} is determined
using the instantaneous steady velocity $\bm v_t^{\rm st,\theta}
\coloneqq-\mu(\bm\nabla U_t-\bm f_t^{\rm nc})
-\theta D\bm\nabla\ln q_t^\theta$ as a solution of
\begin{align}
 \bm\nabla\cdot(q_t^\theta\bm v_t^{\rm st,\theta})&=0.
 \label{eq:another_temperature_reference}
\end{align}
We explain how to construct a decomposition whose relaxation target $\gamma_t$
coincides with this \redtext{intended target}.}
Setting $r_t\coloneqq\rho_t/q_t^\theta$, we find that
$\bm v_t-\bm v_t^{\rm st,\theta}
=D\bm\nabla[(\theta-1)\ln q_t^\theta-\ln r_t]$ is a gradient.
Thus, the condition for $\gamma_t=q_t^\theta$ can be written as
$\bm h_t=\bm v_t^{\rm st,\theta}$.
Therefore, if there exists a smooth function $K_\theta$ satisfying
$K_\theta(1)=0$, $1-(\theta-1)rK_\theta'(r)\geq0$, and
\begin{align}
 \ln q_t^\theta(\bm x)&=K_\theta(r_t(\bm x))+C_t,
 \label{eq:another_temperature_generator}
\end{align}
where $C_t$ is a spatially uniform constant, then the decomposition with
$\gamma_t=q_t^\theta$ is constructed by choosing the generating function
$f=f_{\rm temp}$ given by
\begin{align}
 f_{\rm temp}(r)&=f_{\rm KL}(r)
       -(\theta-1)\int_1^r K_\theta(s)\dd s.
 \label{eq:another_temperature_f_definition}
\end{align}
\par
This construction can be interpreted as evaluating the approach to the steady
state at a desired temperature when the heat-bath temperature changes
uncontrollably.
We denote the excess and housekeeping EPRs in this case by $\dot\Sigma_t^{\rm
temp, ex}$ and $\dot\Sigma_t^{\rm temp, hk}$, respectively.
In contrast to the HS decomposition, which is based on the instantaneous
steady \redtext{state} at the actual bath temperature, $\dot\Sigma_t^{\rm
temp,ex}$ represents the dissipation associated with the approach to the
steady \redtext{state} at the prescribed temperature, whereas
$\dot\Sigma_t^{\rm temp,hk}$ represents the dissipation associated with the
flow that preserves \redtext{the same steady state}.
This decomposition thus evaluates, along the actual process with a varying
temperature, the approach to the steady \redtext{state} at a prescribed
temperature under the same external forces.
Moreover, it decomposes the velocity into the part describing this approach
and the part preserving the steady \redtext{state} chosen as the relaxation
target, while keeping the two mutually orthogonal.
We refer to this decomposition as the temperature-based decomposition.

\subsection{Further remarks}
\subsubsection{Existence of the decomposition}
We now \redtext{discuss a sufficient condition for} the existence of the
decomposition.
The construction requires the existence of the relaxation target $\gamma_t$
defined by \redtext{Eqs.~(\ref{eq:condition_a})--(\ref{eq:condition_b})},
which is not guaranteed for an arbitrary generating function $f$.
Assume that $\Omega$ is a bounded domain with periodic boundary conditions or
with the condition that the probability current vanishes at the boundary. If
the generating function $f$ satisfies
\begin{align}
    \lambda\leq r f''(r)\leq\Lambda
    \qquad (r>0)
    \label{eq:decomposability}
\end{align}
for some positive constants $\lambda,\Lambda$, the decomposition can be
constructed for any $\rho_t,\ \bm v_t$.
Detailed assumptions and the proof are given in
Appendix~\ref{app:general_solvability}.\par
\redtext{A typical example satisfying this condition is the family of scaled
KL divergences, i.e., $f(r)=f_{\rm KL}(r)$ in Eq.~\eqref{eq:mn_scaled_family}.
The HS decomposition is the case for $\epsilon=1$, and the MN decomposition is
obtained as its appropriate scaling limit $\epsilon\to0$.}
Thus, the HS and MN decompositions are naturally identified as a typical
example and a limit, respectively, of the case where the decomposition always
exists for any $\rho_t$ and $\bm v_t$.
Note, however, that condition~(\ref{eq:decomposability}) does not require $r
f''(r)$ to be constant; it also allows non-KL-type generating functions as
long as $r f''(r)$ is positive everywhere and remains between positive finite
values as $r\to0,\infty$.

Moreover, Eq.~(\ref{eq:decomposability}) is only a sufficient condition:
even when it is not satisfied, the decomposition may still be constructed,
depending on the given $\rho_t,\ \bm v_t$. In the next section, we consider
decompositions based on the Hellinger divergence and reverse KL divergence on
a one-dimensional ring and explicitly give the conditions under which each
decomposition exists for given $\rho_t,\ \bm v_t$.

\subsubsection{Formulation by Helmholtz decomposition}
Furthermore, our decomposition can also be viewed as a weighted Helmholtz
decomposition based on an inner product with a different weight.
Let \(w_t\coloneqq\gamma_t^2/\rho_t\) and consider the $w_t$--weighted inner
product. Then, by multiplying both sides of the decomposition $\bm
v_t=\bm\nabla\phi_t+\bm h_t$ by $\rho_t/\gamma_t$ and
using~\eqref{eq:rho_gamma}, we obtain
\begin{align}
  &\frac{\rho_t}{\gamma_t}\bm v_t
  =
  \bm\nabla\psi_t
  +
  \frac{\rho_t}{\gamma_t}\bm h_t,
  \notag\\
  &\bm\nabla\cdot
  \left[
  w_t\frac{\rho_t}{\gamma_t}\bm h_t
  \right]
  =
  0,\\
  &\left\langle
  \bm\nabla\psi_t,
  \frac{\rho_t}{\gamma_t}\bm h_t
  \right\rangle_{w_t}
  =
  0.
  \label{eq:f_weighted_helmholtz}
\end{align}
Therefore, the decomposition of the scaled velocity \(\rho_t\bm v_t/\gamma_t\)
can be regarded as a Helmholtz decomposition with respect to the
$w_t$--weighted inner product into a gradient component and a component with
vanishing weighted divergence.\par
In this representation, Eq.~(\ref{eq:condition_b}) can also be understood as a
projection condition that determines the relaxation target $\gamma_t$
self-consistently.
Denoting by $\mathsf P_w$ the orthogonal projection onto the gradient
component of the Helmholtz decomposition with respect to a positive weight
$w$, we have $\mathsf P_{w_t}(r_t\bm v_t) =
D\bm\nabla\bigl[f(r_t)-r_tf'(r_t)\bigr]$ for $r_t=\rho_t/\gamma_t$ and
$w_t=\gamma_t^2/\rho_t$.
Therefore, Eq.~(\ref{eq:condition_b}) is equivalent to
\begin{align}
  \mathsf P_{w_t}(r_t\bm v_t)
  =
  \bm\nabla\psi_t.
  \label{eq:f_helmholtz_consistency}
\end{align}
The left-hand side is the gradient component determined by the Helmholtz
decomposition with weight $w_t$, whereas the right-hand side is the
gradient-flow velocity of the $f$-divergence for the same $\gamma_t$,
multiplied by $\rho_t/\gamma_t$.
Our construction can thus be regarded as simultaneously determining the
relaxation target $\gamma_t$ and the decomposition so that these two sides
coincide.
\parhead{Examples}\label{sec:examples}
We illustrate the above decomposition with dynamics on a one-dimensional ring
and perform numerical calculations.
The state space $\Omega$ is the ring $[0,L)$ with period \(L=2\pi\) with
periodic boundary conditions.
We use dimensionless units, with $\mu=D=1$ in the first two settings.
We perform numerical calculations in three settings.
In the first setting, we compare the four cases obtained by constructing our
decomposition with four different $f$-divergences for the same dynamics.
In the second setting, we consider the equilibrium-based decomposition
introduced in Sec.~\ref{subsubsec:another_equilibrium}.
In the third setting, we consider the temperature-based decomposition
introduced in Sec.~\ref{subsubsec:another_temperature}.
\subsection{Four divergences}

We consider the case where a uniform nonconservative force $f^{\rm
nc}(x)=F\mathrel{>}0$ is applied under the potential $U(x)=-A\cos x$.
In this example, we decompose the same \((\rho_t,v_t)\) obtained at each time
in four ways: the MN limit, \(f_{\rm KL}(r)\coloneqq r\ln r-r+1\)
corresponding to the KL divergence (HS decomposition), \(f_{\rm
RKL}(r)\coloneqq -\ln r+r-1\) corresponding to the reverse KL divergence, and
\(f_{\rm H}(r)\coloneqq (\sqrt r-1)^2\) corresponding to the Hellinger
divergence.

For the reverse KL and Hellinger divergences, the existence of the
decomposition is not automatically guaranteed.
To state the condition for solvability of Eq.~\eqref{eq:condition_b}, we
define the quantities
\begin{align}
 \mathcal S_t(x)
 &\coloneqq 
 Fx+A\cos x-\ln\rho_t(x),\label{eq:ring_example_S}\\
 \Delta_t(x)
 &\coloneqq 
 \mathcal S_t(x)-
 \min_{0\leq \ell\leq L}\mathcal S_t(x+\ell).
\label{eq:ring_example_delta}
\end{align}
Here, \(\partial_x\mathcal S_t=v_t\), and \(\Delta_t(x)\geq0\) represents the
maximum decrease of \(\mathcal S_t\) along one loop around the ring starting
from \(x\).
Then, the necessary and sufficient condition for the existence of a solution
is given by
\begin{align}
 \int_0^L
 \rho_t(x)\Delta_t(x)^p\,\dd x
 &<
 1,
 &
 p&=
 \begin{cases}
 1,&\text{reverse KL},\\
 2,&\text{Hellinger}
 \end{cases}.
\label{eq:ring_example_solvability}
\end{align}
The derivation of Eq.~\eqref{eq:ring_example_solvability} is given in
Appendix~\ref{sec:supp_ring}.
In the numerical calculations, the initial \redtext{state} is
\(\rho_0(x)\propto\exp[\kappa_0\cos(x-x_0)]\), and the parameters are
\(L=2\pi\), \(D=1\), \(F=1\), \(A=1.5\), \(\kappa_0=1.5\), and \(x_0=2\).
Since the above condition is satisfied at all output times in $0\leq t\leq3$,
the EPR can also be decomposed according to our construction for the reverse
KL and Hellinger divergences.\par
\begin{figure}[t]
 \centering
 \includegraphics[width=230pt]
 {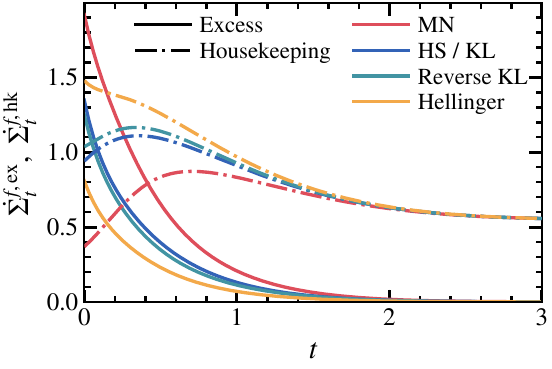}
 \caption{
Time evolution of the excess and housekeeping
EPRs for the same process. The parameters are
\(L=2\pi\), \(D=1\), \(F=1\), and \(A=1.5\).
The initial \redtext{state} is
\(\rho_0(x)\propto
\exp[\kappa_0\cos(x-x_0)]\), with
\(\kappa_0=1.5\) and \(x_0=2\). Solid lines denote \(\dot\Sigma_t^{f,\rm ex}\),
and dash-dotted lines denote \(\dot\Sigma_t^{f,\rm hk}\). Red, blue, teal, and
orange correspond to MN, HS/KL, reverse KL, and Hellinger, respectively.
Different choices of the generating function or its scaling limit give
different decompositions.
 }\label{fig:ring_example}
\end{figure}
Figure~\ref{fig:ring_example} compares the four excess--housekeeping
decompositions for the same dynamics \(\rho_t,v_t\).
Since the sum of the solid and dash-dotted lines of each color equals the
total EPR, the differences between the curves represent differences in how the
same EPR is decomposed into the excess and housekeeping parts.
The numerical results also confirm that the MN decomposition gives the largest
excess part and the smallest housekeeping part.
In this example, since the drift is time independent, all the excess parts
decay to zero at long times, and the housekeeping parts approach a common
finite value that sustains the steady current driven by the nonconservative
force.

\subsection{Equilibrium-based decomposition}
\redtext{We consider the equilibrium--based decomposition for a uniform
potential $U_t(x)=0$.}
\redtext{We take positive parametors $\lambda,\ \omega,\ c$ and
$\varepsilon\in (0,1)$, and set $a_c(t)\coloneqq\varepsilon
e^{-(1+c\lambda)t}$ and $\xi_c(x,t)\coloneqq x-x_0-c\omega t$.
We consider the nonconservative force given by}
\begin{align}
 f_t^{{\rm nc},(c)}(x)
 &=c\left[
 \omega+\frac{\lambda a_c(t)\sin\xi_c(x,t)}
 {1+a_c(t)\cos\xi_c(x,t)}
 \right].
 \label{eq:ring_eq_only_protocol}
\end{align}
\redtext{The parameter $c$ thus sets the strength of the applied
nonconservative force.} The solution starting from the common initial
\redtext{state} $\rho_0(x)=[1+\varepsilon\cos(x-x_0)]/L$ is
\begin{align}
 \rho_t^{(c)}(x)
 &=\frac{1+a_c(t)\cos\xi_c(x,t)}{L},
 \notag\\
 v_t^{(c)}(x)
 &=c\omega-(1+c\lambda)\partial_x\ln\rho_t^{(c)}(x).
 \label{eq:ring_eq_only_solution}
\end{align}
We consider two processes with $c=2$ and $4$, and for each $c$ decompose the
same $(\rho_t^{(c)},v_t^{(c)})$ in two ways: by the HS decomposition and by
the equilibrium-based decomposition.
Although different values of $c$ correspond to different processes, the two
decompositions are compared under a common protocol for each $c$.
\par

The equilibrium \redtext{state} of the system without the nonconservative
force is $q=1/L$.
Choosing the generating function as $f^{(c)}=(1+c\lambda)f_{\rm KL}$, we find
that $\gamma_t=q$, $\partial_x\phi_t=-(1+c\lambda)\partial_x\ln\rho_t^{(c)}$,
and $h_t=c\omega$ satisfy the
conditions~\redtext{\eqref{eq:condition_a}--\eqref{eq:condition_b}} of our
construction.
The corresponding excess and housekeeping EPRs are obtained as
\begin{align}
 \dot\Sigma_t^{{\rm eq},\rm ex}
 &=(1+c\lambda)^2
 \left[1-\sqrt{1-a_c(t)^2}\right],
 \notag\\
 \dot\Sigma_t^{{\rm eq},\rm hk}
 &=c^2\omega^2.
 \label{eq:ring_eq_only_rates}
\end{align}
On the other hand, the HS decomposition uses $f_{\rm KL}$ and takes as its
relaxation target the steady \redtext{state} of the system with the
nonconservative force fixed at each time.
In the numerical calculations, we set the parameters to $\varepsilon=0.8$,
$x_0=2$, $\lambda=0.5$, and $\omega=0.35$, and perform the calculations for
$c=2$ and $c=4$.
\par

\begin{figure*}[t]
 \centering
 \includegraphics[width=470pt]{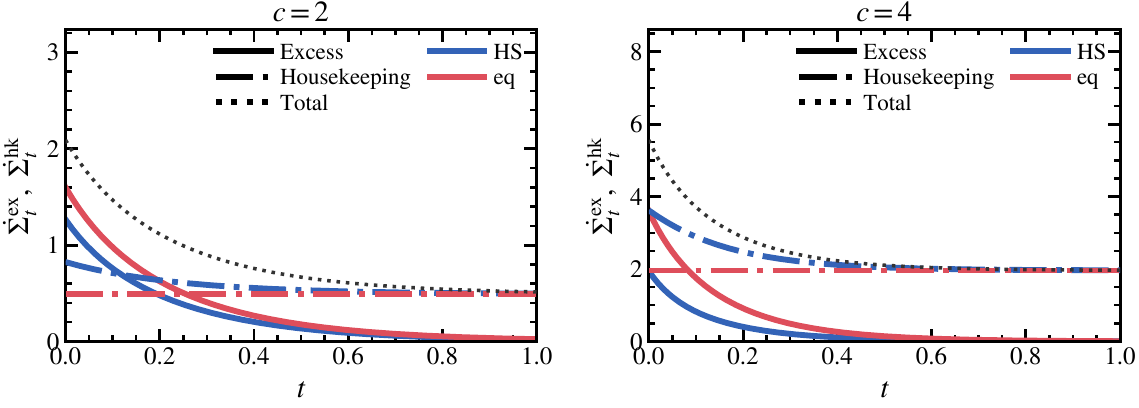}
 \caption{
 Time evolution of the excess and housekeeping EPRs for the processes with
$c=2,\ 4$.
 The parameters are $L=2\pi$, $D=\mu=1$,
 $\varepsilon=0.8$, $x_0=2$, $\lambda=0.5$, and $\omega=0.35$,
 and the initial \redtext{state} is $\rho_0(x)=[1+\varepsilon\cos(x-x_0)]/L$.
 Solid and dash-dotted lines denote the excess and housekeeping EPRs,
respectively,
 with blue and red corresponding to the HS and eq decompositions, respectively.
 Black dotted lines denote the total EPR, which is common to both
decompositions.
 }
 \label{fig:ring_equilibrium_c2}
\end{figure*}

Figure~\ref{fig:ring_equilibrium_c2} compares the two decompositions for the
processes with $c=2$ and $c=4$.
In each panel, the sum of the solid and dash-dotted lines of the same color
equals the total EPR, so the differences between the curves represent
differences in how the same EPR is decomposed into the excess and housekeeping
parts.
While the HS decomposition refers to the instantaneous steady \redtext{state}
of the driven system, the equilibrium-based decomposition decomposes the EPR
into a part associated with the approach to the uniform equilibrium
\redtext{state} and a part associated with the flow that preserves this
equilibrium \redtext{state}.
The difference between the equilibrium-based and HS decompositions is larger
for $c=4$, with the larger coefficient of the nonconservative force, than for
$c=2$.
\subsection{Temperature-based decomposition}
\redtext{For the actual bath temperature $T$, we consider the
temperature-based decomposition with respect to the process in which the
temperature is scaled to $T_{\rm ref}=\theta T$ for $0<\theta<2$.
Recall that $\theta=1$ recovers the Hatano--Sasa decomposition.
}
Setting $\nu_\theta=\mu(1/\theta-1/2)$, we define $U_t(x)=-\tfrac12\ln p_t(x)$
and \redtext{$f_t^{\rm nc}(x)=\omega/\sqrt{p_t(x)}$} by using the function
$p_t$ satisfying
\begin{align}
 \partial_t p_t
 &=\nu_\theta\partial_x^2p_t-\mu\omega\partial_x\sqrt{p_t},
 &p_0(x)&=1+\varepsilon\cos(x-x_0).
 \label{eq:temperature_example_p}
\end{align}
\redtext{In this numerical calculation, we choose units in which
$k_\mathrm{B}T_\mathrm{ref}=1$, so that $D=\mu/\theta$.}
\par
\redtext{We apply the construction in Sec.~\ref{subsubsec:another_temperature}
to this protocol.
The solution for the Fokker--Planck equation is given by $\rho_t=p_t/L$.
At the reference temperature $T_{\rm ref}=\theta T$ under the same external
forces, the instantaneous steady state is
$q_t^\theta=\sqrt{p_t}/\int_0^L\sqrt{p_t}\,dx$, and the corresponding
steady-state velocity $v_t^{\rm st,\theta}=\mu\omega/\sqrt{p_t}$ satisfies
Eq.~\eqref{eq:another_temperature_reference}.
Setting $r_t=\rho_t/q_t^\theta$, we have $\ln q_t^\theta=\ln r_t+C_t$, where
$C_t$ is spatially uniform.
Thus, Eq.~\eqref{eq:another_temperature_generator} holds with $K_\theta(r)=\ln
r$, which is smooth for $r>0$ and satisfies $K_\theta(1)=0$.
Substituting this choice into Eq.~\eqref{eq:another_temperature_f_definition}
gives a convex generating function $f_{\rm temp}(r)=(2-\theta)f_{\rm KL}(r)$.
With $\gamma_t=q_t^\theta$, the housekeeping and excess velocities are given
by $h_t^\theta=v_t^{\rm st,\theta}=\mu\omega/\sqrt{p_t}$ and
$u_t^\theta=v_t-h_t^\theta=-\nu_\theta\partial_x\ln p_t$, respectively.}
We set $\mu=10$, $\varepsilon=0.8$, $x_0=2$, and $\omega=0.2$, and for each of
the processes constructed with $\theta=1.2$ and $1.4$, we compare the
temperature-based decomposition and HS decomposition for the same
$\rho_t,v_t,D$ over $0\le t\le1$.
\par
\begin{figure*}[t]
 \centering
 \includegraphics[width=460pt]{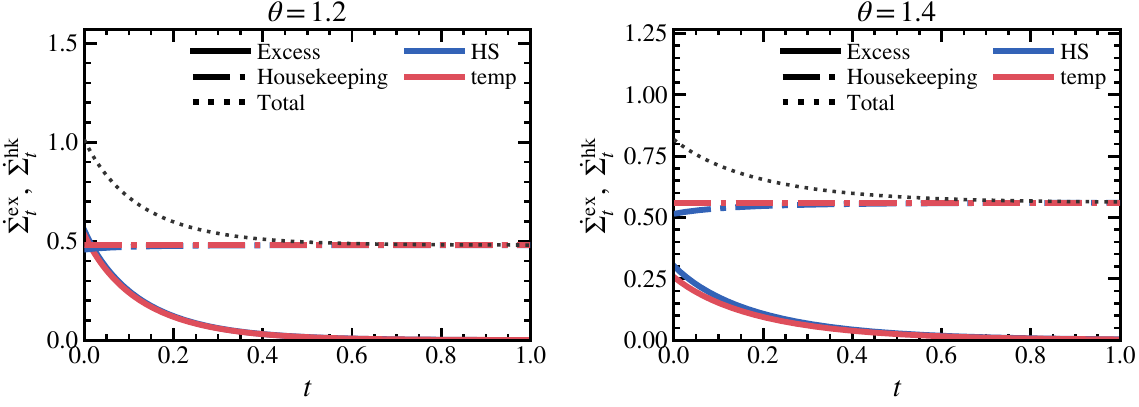}
 \caption{
For the processes with temperature ratios $\theta=1.2,\ 1.4$,
the time evolution of the excess and housekeeping EPRs over $0\le t\le1$ is
shown.
The parameters are $L=2\pi$, $\mu=10$, $D=\mu/\theta$,
$\varepsilon=0.8$, $x_0=2$, and $\omega=0.2$,
and the initial \redtext{state} is $\rho_0(x)=[1+\varepsilon\cos(x-x_0)]/L$.
Solid and dash-dotted lines denote the excess and housekeeping EPRs,
respectively,
with blue and red corresponding to the HS and temp decompositions,
respectively.
Black dotted lines denote the total EPR, which is common to both
decompositions.
}
 \label{fig:num3}
\end{figure*}

Figure~\ref{fig:num3} shows the time evolution of the excess and housekeeping
EPRs. For both decompositions, the sum of the excess and housekeeping parts
equals the same total EPR, and the differences between the curves reflect how
dissipation is decomposed into the two parts. \redtext{For} $\theta=1$, our
decomposition coincides with the HS decomposition. Of the two conditions
considered here, $\theta=1.2$ gives a smaller difference from HS in both parts
than $\theta=1.4$.

\parhead{Discussion}\label{sec:discussion}
In this work, we have constructed a class of decompositions based on general
$f$-divergences into an excess part describing relaxation \redtext{to a
desired target} and a housekeeping part preserving the \redtext{same}
relaxation target.
The construction conditions~\eqref{eq:condition_a},~\eqref{eq:housekeeping_def%
}, and~\eqref{eq:condition_b}
determine the relaxation target $\gamma_t$ and the decomposition of the
velocity. 
\redtext{
This decomposition satisfies the orthogonality with respect to the
$\rho_t$--weighted inner product and the requirement that the housekeeping
velocity vanishes in the absence of nonconservative forces. 
It also recovers the HS decomposition for the KL divergence and the MN
decomposition in an appropriate scaling limit. 
}
These properties support its interpretation as a class of excess--housekeeping
decompositions. \redtext{In our construction,} the excess EPR has two physical
interpretations: the relaxation rate of the
$f$-divergence~\eqref{eq:excess_differential}, and the minimal squared norm
among velocities that induce the same instantaneous time evolution as the
velocity $\bm v_t$ when acting on $\gamma_t$~\eqref{eq:f_minimum_action}. This
unifies the relaxation interpretation of the HS decomposition and the
optimal-transport interpretation of the MN decomposition through the common
structure in which the housekeeping part preserves the same relaxation target.
Specifically, the HS decomposition belongs to the family of scaled KL
divergences satisfying the sufficient condition~\eqref{eq:decomposability} for
the existence of the decomposition, and the MN decomposition arises as a limit
of this family. 
Within this framework, we have also constructed the equilibrium-based
decomposition and temperature-based decomposition as physically meaningful
excess--housekeeping decompositions beyond the HS and MN decompositions. These
quantify the extent of the change toward a desired steady \redtext{state}
when, respectively, the nonconservative force or temperature differs from its
desired value.\par
\par
Finally, we discuss future directions.
First, it is important to clarify the range of $(f,\rho_t,\bm v_t)$ for which
the decomposition exists and the stability of its solution, and to provide
criteria for choosing the generating function $f$ according to the quantities
to be controlled or the relaxation to be observed.
For existing excess--housekeeping decompositions, the estimation of the EPR
from trajectory data via variational
representations~\cite{DechantSasaIto2022PRR} and thermodynamic uncertainty
relations for the excess and housekeeping parts~\cite{KamijimaEtAl2023} have
been studied.
Connecting our representation~\eqref{eq:f_minimum_action} to these results and
investigating inequalities that bound dissipation \redtext{for} the
equilibrium-based and temperature-based decompositions are also important
steps toward concrete applications.
Furthermore, formulating our framework for discrete systems is an important
direction for clarifying its relation to existing geometric decompositions for
Markov jump processes and chemical reaction
networks~\cite{Kobayashi2022HessianDecomposition,YoshimuraEtAl2023,Kolchinsky2%
026GeneralizedFreeEnergy,NagayamaEtAl2025,SagawaHayakawa2011}.

\begin{acknowledgments}
We thank Kaito Tojo for valuable discussions. This work is supported by JST
ERATO Grant No.~JPMJER2302, Japan. R.N. is supported by the World-leading
Innovative Graduate Study Program for Materials Research, Information, and
Technology (MERIT-WINGS) of the University of Tokyo and by JSPS KAKENHI Grant
No.~JP26KJ0820. 
\redtext{S.S. is supported by JSPS KAKENHI Grant Number JP26K17047. }
T.S. is supported by JST CREST Grant No.~JPMJCR20C1 and by the Institute of AI
and Beyond of the University of Tokyo.
\end{acknowledgments}

\begin{appendix}

\section{Recovery of the MN decomposition}
\label{app:mn_limit}

In this section, we show that the decomposition corresponding to
$f_\epsilon$ reduces to the MN decomposition in the limit $\epsilon\to 0$
for the family in Eq.~\eqref{eq:mn_scaled_family}. 
We {show the existence of} the small-$\epsilon$ solution {by applying the
implicit function theorem.
The smooth dependence of this solution on $\epsilon$ then gives estimates of
the errors in the velocity and EPR.}

Throughout this section, we fix the time $t$ and omit the subscript $t$.
{Here, we assume that $\Omega$ is connected and, in the
nonperiodic case, bounded with a smooth boundary.}
We assume that the probability density $\rho$ {has a positive
infimum throughout the closure of $\Omega$, denoted by $\bar{\Omega}$},
and that $\rho$ and the velocity field $\boldsymbol{v}$ are sufficiently
smooth. If a boundary is present, we impose
$\boldsymbol{v}\cdot\boldsymbol{n}=0$
and $\partial_n\phi=0$ {on $\partial\Omega$}, where
$\boldsymbol{n}$ is the outward unit normal. We also assume that the
generating function $f$ is sufficiently smooth near $r=1$ and satisfies
$f'(1)=0$ and $a:=f''(1)>0$.

We first express the decomposition conditions in terms of the potential.
Let $g$ denote the local inverse of $f'$ near $r=1$; then $g(0)=1$ and
$g'(0)=1/a$. Equation~\eqref{eq:condition_a} is equivalent to
$\rho/\gamma^\epsilon=g(-\epsilon\phi^\epsilon/D)$,
{for an appropriate potential $\phi^\epsilon$. To find a
solution $\phi^\epsilon$ satisfying
Eqs.~\eqref{eq:condition_a}--\eqref{eq:condition_b},
we introduce}
\begin{align}
  \Gamma_\epsilon[\phi]
    &:= \frac{\rho}{g(-\epsilon\phi/D)} \notag\\
    &\mathrel{{=}} \rho+\frac{\epsilon}{Da}\rho\phi+O(\epsilon^2)
 \label{eq:app_mn_gamma_expansion}
\end{align}
for sufficiently small $\epsilon$.
{The relaxation target is then given by
$\gamma^\epsilon=\Gamma_\epsilon[\phi^\epsilon]$, and $\phi^\epsilon$
satisfies}
conservation of the housekeeping current and normalization:
\begin{align}
  \boldsymbol{\nabla}\cdot
  [\Gamma_\epsilon[\phi^\epsilon](\boldsymbol{\nabla}\phi^\epsilon-\boldsymbol%
{v})]
    &= 0, \label{eq:app_mn_map_current}\\
  \int_\Omega \Gamma_\epsilon[\phi^\epsilon]\mathrm{d}\boldsymbol{x}
    &= 1. \label{eq:app_mn_map_normalization}
\end{align}

{Below, we use the implicit function theorem~\cite{Deimling1985} to show the
existence of the solution $\phi^{\epsilon}$ which smoothly converges to
$\phi^{\mathrm{MN}}$ as $\epsilon \to 0$.
For that purpose, we rewrite Eqs.~\eqref{eq:app_mn_map_current} and
\eqref{eq:app_mn_map_normalization} as
\begin{align}
 F(\epsilon,\phi^\epsilon)&=\boldsymbol{0},
 \label{eq:app_mn_implicit_equation}\\
 F(\epsilon,\phi)&\coloneqq
 \begin{pmatrix}
 \boldsymbol{\nabla}\cdot[\Gamma_\epsilon[\phi](\boldsymbol{\nabla}\phi-%
\boldsymbol{v})]\\[1ex]
 \dfrac{Da}{\epsilon}\left(\displaystyle\int_\Omega\Gamma_\epsilon[\phi]%
\mathrm{d}\boldsymbol{x}-1\right)
 \end{pmatrix}.
 \label{eq:app_mn_implicit_map}
\end{align}
The expansion of $\Gamma_\epsilon[\phi]$ shows that the second component of
$F$ tends to $\int_\Omega\rho\phi\,\mathrm{d}\boldsymbol{x}$ as $\epsilon\to0$.
Thus, $F$ extends smoothly to $\epsilon=0$, and $F(0,\phi)=\boldsymbol{0}$ is
equivalent to Eq.~\eqref{eq:orth_MN}.
In particular, $F(0,\phi^{\mathrm{MN}})=\boldsymbol{0}$.}

{To apply the implicit function theorem to the function $F$ at $(\epsilon,
\phi) = (0, \phi^{\mathrm{MN}})$, it remains to show that the derivative of
$F$ with respect to $\phi$ is invertible at this point.
Since $\epsilon = 0$ at this point, the derivative is obtained as
\begin{align}
 &\mathcal{D}F(0,\phi^{\mathrm{MN}})[\psi]
 \notag\\
 &\quad\coloneqq\lim_{\eta\to0}
 \frac{F(0,\phi^{\mathrm{MN}}+\eta\psi)-F(0,\phi^{\mathrm{MN}})}{\eta}
 \notag\\
 &\quad=\begin{pmatrix}
 \boldsymbol{\nabla}\cdot[\rho\boldsymbol{\nabla}\psi]\\[1ex]
 \displaystyle\int_\Omega\rho\psi\mathrm{d}{\bm x}
 \end{pmatrix}
 \quad(\forall\psi).
 \label{eq:app_mn_linearization}
\end{align}
Under the prescribed boundary conditions, the first component of $F$
integrates to zero over $\Omega$.
To invert the derivative, we therefore take any sufficiently smooth function
$h$ satisfying $\int_\Omega h\,\mathrm{d}\boldsymbol{x}=0$ and any number $c
\in \mathbb{R}$, and seek a potential $\psi$ such that
\begin{align}
  \boldsymbol{\nabla}\cdot(\rho\boldsymbol{\nabla}\psi) &= h,
  \label{eq:app_mn_linear_poisson} \\
  \int_\Omega\rho\psi\,\mathrm{d}\boldsymbol{x} &= c.
  \label{eq:app_mn_linear_normalization}
\end{align}
Under the assumptions of this section, the first equation (the weighted
Poisson equation) is known to have a solution, unique up to an additive
constant~\cite{Kono2024,PavliotisStuart2008}.
The second equation then determines this constant.
Thus, Eqs.~\eqref{eq:app_mn_linear_poisson} and
\eqref{eq:app_mn_linear_normalization} determine $\psi$ uniquely for every
such pair $(h,c)$, implying that the derivative $\mathcal{D}F(0,\phi)$ is
invertible.

At this stage, we can apply the implicit function theorem to $F$ at
$(0,\phi^{\mathrm{MN}})$.
For sufficiently small $\epsilon$, it gives a smooth solution $\epsilon
\mapsto \phi^\epsilon$ satisfying $F(\epsilon,\phi^\epsilon)=\boldsymbol{0}$
and  $\phi^0 = \phi^{\mathrm{MN}}$.
Expanding $\phi^\epsilon$ and its spatial gradient about $\epsilon=0$, we
obtain
\begin{align}
 \|\phi^\epsilon-\phi^{\mathrm{MN}}\|_\rho&=O(\epsilon),
 \label{eq:app_mn_potential_error}\\
 \|\boldsymbol{\nabla}\phi^\epsilon-\boldsymbol{\nabla}\phi^{\mathrm{MN}}\|_%
\rho&=O(\epsilon),
 \label{eq:app_mn_gradient_error}
\end{align}
and therefore, by Eq.~\eqref{eq:mn_excess_upper_bound},
\begin{align}
 &\dot{\Sigma}^{\mathrm{MN,ex}}-\dot{\Sigma}^{f_\epsilon,\mathrm{ex}}
 \notag\\
 &\quad=\|\boldsymbol{\nabla}\phi^\epsilon-\boldsymbol{\nabla}\phi^{\mathrm{MN%
}}\|_\rho^2
 \notag\\
 &\quad=O(\epsilon^2).
 \label{eq:app_mn_epr_error}
\end{align}
}

\section{Conditions for realizing a prescribed relaxation target}
\label{app:prescribed_reference}

In this appendix, we present the conditions under which the equilibrium-based
and temperature-based decompositions can be constructed. We fix the time $t$
and omit the subscript $t$. We use the same domain and assumptions on $\rho$
as in Appendix~\ref{app:mn_limit}. In each case, $I$ denotes the range of the
density ratio \redtext{$r=\rho/q$} over the domain, \redtext{where $q$ denotes
a desired target}.

\subsection{Equilibrium-based decomposition}
\label{app:equilibrium_reference}

We use the quantities introduced in Sec.~\ref{subsubsec:another_equilibrium}
of the main text:
the equilibrium density $q$, the density ratio $r=\rho/q$, and
the solution $\chi$ of Eq.~\eqref{eq:another_eq_poisson} with zero
$q$-weighted mean.
If a boundary is present, the boundary condition is
$\partial_n\chi=\mu\bm f^{\rm nc}\cdot\bm n$.
With the prescribed boundary conditions, the Poisson equation and the
normalization fix $\chi$ uniquely.
Using $D=\mu/\beta$, we obtain
\begin{align}
 \bm v&=\mu\bm f^{\rm nc}-D\bm\nabla\ln r,
 \label{eq:app_eq_projection_velocity}\\
 \bm u_q&=\bm\nabla\chi-D\bm\nabla\ln r,
 \label{eq:app_eq_projection_excess}\\
 \bm h_q&=\mu\bm f^{\rm nc}-\bm\nabla\chi,
 \label{eq:app_eq_projection_housekeeping}
\end{align}
with $\bm\nabla\cdot(q\bm h_q)=0$.

For any fixed positive relaxation target $\gamma$, the condition
$\bm\nabla\cdot[\gamma(\bm v-\bm\nabla\phi)]=0$ determines the excess velocity
\redtext{$\bm\nabla\phi$} uniquely. Indeed, if $\zeta$ is the difference
between two candidate potentials, integration by parts with periodic or
no-flux boundary conditions gives $\int_\Omega\gamma|\bm\nabla\zeta|^2\dd\bm
x=0$. Thus, any decomposition realizing $\gamma=q$ must have the velocities in
Eqs.~\eqref{eq:app_eq_projection_excess} and
\eqref{eq:app_eq_projection_housekeeping}.

The remaining question is whether a generating function $f$ can reproduce this
excess velocity, i.e., whether $-D\bm\nabla f'(r)=\bm u_q$.
Using $K_{\rm eq}$ introduced in the main text, this condition is equivalent to
\begin{align}
 \chi&=D K_{\rm eq}(r)+C,
 \label{eq:app_eq_profile_potential}\\
 K_{\rm eq}(s)&=\ln s-f'(s),
 \label{eq:app_eq_profile_generator}
\end{align}
where $C$ is a spatially uniform constant.
Equation~\eqref{eq:app_eq_profile_potential} implies that $\chi$ must depend
on position only through $r$.

Here, $I$ is an interval with a positive minimum and a finite maximum, and
$\int_\Omega q r\dd\bm x=1$ implies $1\in I$.
A suitable $f$ exists if and only if Eqs.~\eqref{eq:app_eq_profile_potential}
and \eqref{eq:app_eq_profile_generator} hold with $K_{\rm eq}(1)=0$ and the
resulting $f$ is convex.
Over the range $I$, the convexity condition reads
\begin{align}
 1-sK_{\rm eq}'(s)\geq0\qquad(s\in I).
 \label{eq:app_eq_convexity}
\end{align}
When these conditions hold, the generating function in
Eq.~\eqref{eq:another_eq_f_definition} satisfies, on $I$,
\begin{align}
 f(s)&=f_{\rm eq}(s),
 \label{eq:app_eq_generator_proof_generator}\\
 f''(s)&=\frac{1-sK_{\rm eq}'(s)}{s}\geq0.
 \label{eq:app_eq_generator_proof_convexity}
\end{align}
Hence, $\bm\nabla\phi=\bm u_q$ holds, whereas $\bm h_q$ preserves $q$.
Therefore, both construction conditions are satisfied with $\gamma=q$.

Only the values of $f$ on the actual range $I$ enter the construction.
To define $f$ for all $s>0$, we extend $sf''(s)$ as a constant beyond each
endpoint of $I$ and integrate $f''$ with $f'(1)=f(1)=0$.
This extension leaves $f$ unchanged on $I$ and keeps $f''$ continuous and
nonnegative, thus preserving convexity.

\subsection{Temperature-based decomposition}
\label{app:temperature_reference}

We use the quantities introduced in Sec.~\ref{subsubsec:another_temperature}
of the main text:
the temperature ratio $\theta>0$ with $\theta\neq1$, steady density
$q^\theta$, density ratio $r=\rho/q^\theta$, and steady velocity $\bm v^{\rm
st,\theta}$.
Under the assumptions \redtext{given at the beginning of this section}, the
steady density in Eq.~\eqref{eq:another_temperature_reference} uniquely exists.

From Eq.~\eqref{eq:velocity}, we obtain
\begin{align}
 \bm v-\bm v^{\rm st,\theta}
 &=D\bm\nabla\bigl[(\theta-1)\ln q^\theta-\ln r\bigr],
 \label{eq:app_temperature_gradient}
\end{align}
where $\bm v^{\rm st,\theta}$ preserves the relaxation target $q^\theta$.
Since the difference between the actual velocity and $\bm v^{\rm st,\theta}$
is a gradient, the uniqueness established above implies that choosing
$\gamma=q^\theta$ fixes the housekeeping velocity to be $\bm h=\bm v^{\rm
st,\theta}$.
Hence, the excess velocity coincides with the right-hand side of
Eq.~\eqref{eq:app_temperature_gradient}.

\redtext{For this excess velocity to be expressed as $-D\bm\nabla f'(r)$, $\ln
q^\theta$ must be a function of $r$ \redtext{alone}.
For given $\rho,\bm v,\theta$, and $q^\theta$, a convex generating function
$f$ that yields a decomposition with the relaxation target $\gamma=q^\theta$
exists if and only if there exist a continuously differentiable function
$K_\theta$ with $K_\theta(1)=0$ and a spatially uniform constant $C$ that
satisfy the following relations:
\begin{align}
 \ln q^\theta&=K_\theta(r)+C,
 \label{eq:app_temperature_profile_density}\\
 f'(s)&=f_{\rm temp}'(s)
 \notag\\
 &=\ln s-(\theta-1)K_\theta(s),
 \label{eq:app_temperature_profile_generator}\\
 1-(\theta-1)sK_\theta'(s)&\geq0
 \qquad(s\in I).
 \label{eq:app_temperature_profile_convexity}
\end{align}
Equation~\eqref{eq:app_temperature_profile_density} requires that $\ln
q^\theta$ depend on position only through $r$.
Equation~\eqref{eq:app_temperature_profile_generator} ensures that
$-D\bm\nabla f'(r)=\bm v-\bm v^{\rm st,\theta}$ and, together with $f(1)=0$,
yields the generating function in
Eq.~\eqref{eq:another_temperature_f_definition}.
The inequality in Eq.~\eqref{eq:app_temperature_profile_convexity} is
equivalent to $f''(s)=[1-(\theta-1)sK_\theta'(s)]/s\geq0$ and thus ensures the
convexity of $f$.
The extension of $f$ outside $I$ is carried out in the same way as for the
equilibrium-based decomposition.}

\section{Proof of a sufficient condition for the existence of the
decomposition}
\label{app:general_solvability}

In this section, we show that, when condition~\eqref{eq:decomposability} is
satisfied, there exists a relaxation target $\gamma_t$ that satisfies the
construction conditions~\redtext{\eqref{eq:condition_a}--\eqref{eq:condition_b%
}}.
Throughout this section, we fix the time $t$ and omit the subscript $t$.
\redtext{Here, we assume that $\Omega$ is connected and, in the nonperiodic
case, bounded with a smooth boundary.}
We assume that the density $\rho$ and the velocity field $\bm v$ are
sufficiently smooth, and that $\rho$ satisfies
$0<\rho_-\leq\rho\leq\rho_+<\infty$ \redtext{for some constants $\rho_-$ and
$\rho_+$}.
If a boundary is present, we impose $\bm v\cdot\bm n=0$ \redtext{on
$\partial\Omega$}.

For the generating function $f$, we assume that $f''$ is continuous and that
Eq.~\eqref{eq:decomposability} holds.
In this section, we write $a(r)\coloneqq r f''(r)$.
With the constants $\lambda,\Lambda$ introduced in the main text, we then have
the bound
\begin{align}
 0<\lambda\leq a(r)\leq\Lambda<\infty
 \qquad(r>0).
 \label{eq:app_uniform_ellipticity}
\end{align}
This condition implies that the effective diffusion coefficient appearing
below is bounded from above and bounded away from zero.

We first rewrite the construction conditions as a steady diffusion equation
for $\gamma$.
Using the density ratio $r$ and condition~\eqref{eq:condition_a} in the main
text, we have
\begin{align}
 \gamma\bm h
 &=\gamma\bm v+D\gamma\bm\nabla f'(\rho/\gamma)
 \notag\\
 &=\left[\bm v+D a(\rho/\gamma)\bm\nabla\ln\rho\right]\gamma
   -D a(\rho/\gamma)\bm\nabla\gamma.
 \label{eq:app_solvability_flux}
\end{align}

Thus, defining
$A_\gamma\coloneqq D a(\rho/\gamma)$ and
$\bm b_\gamma\coloneqq\bm v+A_\gamma\bm\nabla\ln\rho$,
we can transform the condition \eqref{eq:condition_b} into
\begin{align}
 \bm\nabla\cdot
 (A_\gamma\bm\nabla\gamma-\bm b_\gamma\gamma)&=0,
 \label{eq:app_solvability_quasilinear_equation}\\
 \int_\Omega\gamma\dd\bm x&=1.
 \label{eq:app_solvability_quasilinear_normalization}
\end{align}
If a boundary is present, we impose
$(A_\gamma\bm\nabla\gamma-\bm b_\gamma\gamma)\cdot\bm n=0$ \redtext{on
$\partial\Omega$. If there exists $\gamma$ satisfying
Eqs.~\eqref{eq:app_solvability_quasilinear_equation} and
\eqref{eq:app_solvability_quasilinear_normalization}, then $\phi$ and $\bm h$
satisfying Eqs.~\eqref{eq:condition_a} and \eqref{eq:housekeeping_def} also
exist.
In what follows, we therefore consider the conditions for the existence of a
solution of Eqs.~\eqref{eq:app_solvability_quasilinear_equation} and
\eqref{eq:app_solvability_quasilinear_normalization}.}

\redtext{To prove the existence of a solution of
Eqs.~\eqref{eq:app_solvability_quasilinear_equation} and
\eqref{eq:app_solvability_quasilinear_normalization} by Schauder's fixed-point
theorem~\cite{Deimling1985}, we first fix a positive continuous normalized
density $\eta$, with periodicity when required, and solve the following linear
steady equation for $u$:}
\begin{align}
 &\bm\nabla\cdot(A_\eta\bm\nabla u-\bm b_\eta u)=0,
 \label{eq:app_frozen_equation_equation}\\
 &\int_\Omega u\dd\bm x=1
 \label{eq:app_frozen_equation_normalization}
\end{align}
with periodic or no-flux boundary conditions.
By Eq.~\eqref{eq:app_uniform_ellipticity}, the effective diffusion coefficient
is bounded as
\begin{align}
 D\lambda\leq A_\eta&\leq D\Lambda,
 \label{eq:app_frozen_bounds}
\end{align}
whereas $\bm b_\eta$ is bounded because both $\bm v$ and $\bm\nabla\ln\rho$
are bounded.
\redtext{These bounds ensure a unique positive normalized steady density
$u$~\cite{DroniouVazquez2009,Lejay2001} satisfying
Eqs.~\eqref{eq:app_frozen_equation_equation} and
\eqref{eq:app_frozen_equation_normalization}. Moreover, the estimates in
Ref.~\cite{GilbargTrudinger2001} yield
\begin{align}
 0<m_0&\le u(\bm x)\le M_0,
 \label{eq:app_frozen_estimates_pointwise}\\
 |u(\bm x)-u(\bm y)|
 &\le C|\bm x-\bm y|^\alpha,
 \label{eq:app_frozen_estimates_holder}\\
 \|\bm\nabla u\|_{L^2(\Omega)}&\le C.
 \label{eq:app_frozen_estimates_gradient}
\end{align}
Here, $m_0,M_0,C>0$ and $0<\alpha<1$ are independent of $\eta$, and $\bm x,\bm
y\in\overline\Omega$.

Let $\mathcal K$ be the closed bounded convex set of continuous normalized
densities on $\overline\Omega$ satisfying $m_0\le\eta\le M_0$, with
periodicity when required.
Define $\mathcal T\eta=u$ by Eqs.~\eqref{eq:app_frozen_equation_equation} and
\eqref{eq:app_frozen_equation_normalization}.
The bound \eqref{eq:app_frozen_estimates_pointwise} gives $\mathcal T(\mathcal
K)\subseteq\mathcal K$, and the bound \eqref{eq:app_frozen_estimates_holder}
makes $\mathcal T(\mathcal K)$ relatively compact in the uniform
norm~\cite{Deimling1985}.
Consider a sequence $\eta_n$ in $\mathcal{K}$ such that $\eta_n\to\eta$
uniformly in $\mathcal K$.
Then, continuity of $a$ gives $A_{\eta_n}\to A_\eta$ and $\bm b_{\eta_n}\to\bm
b_\eta$ uniformly.
Combining this coefficient convergence with the gradient bound in
Eq.~\eqref{eq:app_frozen_estimates_gradient}, we find that any uniform
subsequential limit of $\mathcal T\eta_n$ satisfies
Eqs.~\eqref{eq:app_frozen_equation_equation} and
\eqref{eq:app_frozen_equation_normalization} with coefficients $A_\eta$ and
$\bm b_\eta$ and the prescribed boundary conditions.
By uniqueness of the normalized solution of
Eqs.~\eqref{eq:app_frozen_equation_equation} and
\eqref{eq:app_frozen_equation_normalization}, this limit is $\mathcal T\eta$.
The relative compactness of $\mathcal T(\mathcal K)$ therefore implies that
$\mathcal T\eta_n\to\mathcal T\eta$ uniformly.
This proves the continuity of $\mathcal T$.

Schauder's fixed-point theorem~\cite{Deimling1985} therefore ensures existence
of a density $\gamma$ satisfying $\gamma=\mathcal T\gamma$, which solves
Eqs.~\eqref{eq:app_solvability_quasilinear_equation} and
\eqref{eq:app_solvability_quasilinear_normalization}.
}

\section{Solvability on a one-dimensional ring}
\label{sec:supp_ring}

\subsection{Necessary and sufficient condition for solvability}

We use the one-dimensional ring setup in which the four decompositions are
compared in the main text.
In the following proof, we take $F>0$, fix the time $t$, and omit the
subscript $t$.
Let $\rho(x)>0$ be a smooth normalized periodic density.

To treat reverse KL and Hellinger together,
we use Eq.~\eqref{eq:ring_example_solvability} with its parameter $p$ and
define
\begin{align}
 z&\mathrel{\coloneqq}\left(\frac{\gamma}{\rho}\right)^{1/p},
 \label{eq:supp_ring_pz_definition}
\end{align}
so that Eq.~\eqref{eq:condition_a} gives $\phi=z-1$,
and the relaxation target can be written as $\gamma=\rho z^p$.
Equation~\eqref{eq:condition_b} implies that the housekeeping probability
current is constant.
Denoting this current by $\jmath$, we can transform Eq.~\eqref{eq:condition_b}
into the boundary value problem
\begin{align}
 &\partial_x z(x)
 =
 v(x)-\frac{\jmath}{\rho(x)z(x)^p},
 \label{eq:supp_ring_bvp_equation}\\
 &z(x+L)
 =z(x)>0,
 \label{eq:supp_ring_bvp_periodicity}\\
 &\int_0^L\rho(x)z(x)^p\,\dd x
 =1.
 \label{eq:supp_ring_bvp_normalization}
\end{align}
\redtext{Equation \eqref{eq:supp_ring_bvp_equation} corresponds to
Eq.~\eqref{eq:condition_b}.
Equation~\eqref{eq:supp_ring_bvp_periodicity} imposes periodicity and
positivity of $\gamma$, and Eq.~\eqref{eq:supp_ring_bvp_normalization} imposes
its normalization.}

\redtext{Equation~\eqref{eq:velocity} gives $v=F-A\sin x-\partial_x\ln\rho$,
which satisfies
$\int_0^L v(x)\,\dd x=LF>0$. Therefore, integrating
Eq.~\eqref{eq:supp_ring_bvp_equation} over one period shows that
\begin{align}
 \jmath
 &=
 \frac{LF}{
 \displaystyle\int_0^L
 \frac{\dd x}{\rho(x)z(x)^p}}
 >0
 \label{eq:supp_ring_current}
\end{align}
holds.}

\redtext{To study solvability}, we first drop the normalization condition, fix
$\jmath>0$, and find a positive periodic solution of
\redtext{Eq.~\eqref{eq:supp_ring_bvp_equation}.}
We integrate \redtext{Eq.~\eqref{eq:supp_ring_bvp_equation}} backward from
$x=L$ to $x=0$.
\redtext{In the backward integration, the positive term $\jmath/[\rho z^p]$
grows as $z$ decreases, so that $z$ never reaches zero.}
Starting from $z(L)=a>0$, we denote the value after one period by $\mathcal
P_{\jmath}(a)\coloneqq z(0)$.

\redtext{For fixed $\jmath>0$, let $\rho_{\max}=\max_{x\in[0,L]}\rho(x)$ and
choose $a>0$ sufficiently small that
\begin{align}
 \frac{\jmath}{\rho_{\max}a^p}&>\|v\|_\infty.
 \label{eq:supp_ring_small_initial_bound}
\end{align}
Introduce the reverse coordinate $s=L-x$ and write $y(s)=z(L-s)$. Then,
\begin{align}
 y'(s)&=-v(L-s)+\frac{\jmath}{\rho(L-s)y(s)^p},
 \label{eq:supp_ring_backward_equation}\\
 y(0)&=a.
 \label{eq:supp_ring_backward_initial}
\end{align}
At any point where $y(s)=a$, we have
\begin{align}
 y'(s)&\ge-\|v\|_\infty+\frac{\jmath}{\rho_{\max}a^p}>0.
 \label{eq:supp_ring_small_initial_growth}
\end{align}
Thus, $y$ initially moves above $a$. If it returned to $a$ for the first time,
we would have $y'\le0$ at that time, contradicting
Eq.~\eqref{eq:supp_ring_small_initial_growth}. Therefore, for sufficiently
small $a$, we have
\begin{align}
 \mathcal P_{\jmath}(a)&=y(L)>a.
 \label{eq:supp_ring_small_initial_return}
\end{align}
For large $a$,}
\begin{align}
 \lim_{a\to\infty}
 \left[\mathcal P_{\jmath}(a)-a\right]
 &=-\int_0^L v(x)\,\dd x
 \notag\\
 &=-LF<0.
 \label{eq:supp_ring_return_limits}
\end{align}
\redtext{Moreover, the derivative with respect to the initial value $a$
satisfies
\begin{align}
 0<\mathcal P_{\jmath}'(a)
 &=
 \exp\!\left[
 -p\jmath\int_0^L
 \frac{\dd x}{\rho(x)z(x)^{p+1}}
 \right]
 <1,
 \label{eq:supp_ring_return_derivative}
\end{align}
and hence $\mathcal P_{\jmath}(a)-a$ is strictly decreasing in $a$.}
Therefore, $\mathcal P_{\jmath}(a)=a$ has a unique solution $a$, and for each
$\jmath>0$, we obtain a unique positive periodic solution of
Eq.~\eqref{eq:supp_ring_bvp_equation} denoted by  $z_{\jmath}$.

To find the value of the current $\jmath$ satisfying the normalization, we
next examine how the periodic solution depends on the current.
Let $0<\jmath_1<\jmath_2$.
At any point where $z_{\jmath_2}-z_{\jmath_1}$ vanishes, we have
\begin{align}
 \partial_x\left(z_{\jmath_2}-z_{\jmath_1}\right)
 &=-\frac{\jmath_2-\jmath_1}{\rho z_{\jmath_1}^{p}}<0,
 \label{eq:supp_ring_crossing}
\end{align}
\redtext{which shows that every intersection of $z_{\jmath_1}$ and
$z_{\jmath_2}$ is a crossing from $z_{\jmath_2}>z_{\jmath_1}$ to
$z_{\jmath_2}<z_{\jmath_1}$ as $x$ increases.
Periodicity would require any such crossing to be accompanied by a crossing in
the opposite direction.
Equation~\eqref{eq:supp_ring_crossing} excludes crossings in the opposite
direction, so the two periodic solutions cannot cross.}

\redtext{Suppose that $z_{\jmath_2}(x)<z_{\jmath_1}(x)$ at every point. Since
$p>0$ and $\rho>0$, we have
\begin{align}
 \int_0^L\frac{\dd x}{\rho z_{\jmath_2}^{p}}
 &>\int_0^L\frac{\dd x}{\rho z_{\jmath_1}^{p}}.
 \label{eq:supp_ring_reciprocal_order}
\end{align}
Equation~\eqref{eq:supp_ring_current} and $LF>0$ then give
\begin{align}
 \jmath_2
 &=\frac{LF}{\displaystyle\int_0^L\frac{\dd x}{\rho z_{\jmath_2}^{p}}}
 \notag\\
 &<\frac{LF}{\displaystyle\int_0^L\frac{\dd x}{\rho z_{\jmath_1}^{p}}}
 =\jmath_1.
 \label{eq:supp_ring_current_order_contradiction}
\end{align}
This contradicts $\jmath_1<\jmath_2$. Therefore, $z_{\jmath}(x)$ is strictly
increasing in $\jmath$ at every point $x$.
}

\redtext{We next show that there exists a solution satisfying the
normalization condition~\eqref{eq:supp_ring_bvp_normalization}.
We define the mass as
\begin{align}
 M_p(\jmath)
 &\mathrel{\coloneqq}\int_0^L\rho(x)z_{\jmath}(x)^p\,\dd x
 \redtext{.}
 \label{eq:supp_ring_mass}
\end{align}
Equation~\eqref{eq:supp_ring_bvp_normalization} can then be written as
$M_p(\jmath)=1$.
To verify the existence of a solution satisfying this condition, we first
consider the limit of the mass $M_p(\jmath)$ as $\jmath\downarrow0$.}

\redtext{For the function $\mathcal S$ defined in
Eq.~\eqref{eq:ring_example_S} in the main text, we have
\begin{align}
 \partial_x\mathcal S(x)&=v(x).
 \label{eq:supp_ring_running_minimum_derivative}
\end{align}}
\redtext{We use $\Delta$ defined in Eq.~\eqref{eq:ring_example_delta}.}
Integrating \redtext{Eq.~\eqref{eq:supp_ring_bvp_equation}} from $x$ to
$y\in[x,x+L]$, we obtain $z_{\jmath}(x)>\mathcal S(x)-\mathcal S(y)$, and
hence $z_{\jmath}(x)>\Delta(x)$.
Since $z_{\jmath}$ is \redtext{strictly increasing} in $\jmath$, the limit as
$\jmath\downarrow0$ exists.
Suppose \redtext{that, at some point \redtext{$x_*$},
\begin{align}
 a_*&\coloneqq\lim_{\jmath\downarrow0}z_{\jmath}(x_*)>\Delta(x_*)
 \label{eq:supp_ring_zero_current_initial_value}
\end{align}
holds.
Let $z_0$ solve $\partial_x z_0=v$ with $z_0(x_*)=a_*$.
Since $\partial_x\mathcal S=v$, we have
\begin{align}
 z_0(x)&=a_*+\int_{x_*}^{x}v(s)\,\dd s
 \notag\\
 &=a_*+\mathcal S(x)-\mathcal S(x_*).
 \label{eq:supp_ring_zero_current_solution}
\end{align}
By the definition of $\Delta$ in Eq.~\eqref{eq:ring_example_delta},
\begin{align}
 &\min_{x\in[x_*,x_*+L]}z_0(x)
 \notag\\
 &\quad=a_*+\min_{0\leq\ell\leq L}\mathcal S(x_*+\ell)-\mathcal S(x_*)
 \notag\\
 &\quad=a_*-\Delta(x_*)>0.
 \label{eq:supp_ring_zero_current_positive_lower_bound}
\end{align}}
\redtext{
Hence, the solution $z_0$ is positive and avoids the singularity
$z_{\jmath}(x)=0$.
Moreover, $z_0$ solves Eq.~\eqref{eq:supp_ring_bvp_equation} with $\jmath=0$,
and the solution of Eq.~\eqref{eq:supp_ring_bvp_equation} is unique and
depends continuously on $\jmath$.
Therefore, the solutions of Eq.~\eqref{eq:supp_ring_bvp_equation} for small
$\jmath$ would approach this solution $z_0$.

The function $\mathcal S$ satisfies $\mathcal S(x+L)=\mathcal S(x)+LF$.
Substituting this into Eq.~\eqref{eq:supp_ring_zero_current_solution} gives
\begin{align}
 z_0(x+L)&= z_0(x)+LF.
 \label{eq:supp_ring_running_minimum_increment}
\end{align}}
This contradicts the periodicity~\eqref{eq:supp_ring_bvp_periodicity} of the
solution.
We therefore conclude by contradiction that, at every point $x$,
\begin{align}
 \lim_{\jmath\downarrow0}z_{\jmath}(x)
 &=\Delta(x),
 \label{eq:supp_ring_small_current_profile}
\end{align}
which yields the limit of the mass
\begin{align}
 \lim_{\jmath\downarrow0}M_p(\jmath)
 &=\int_0^L\rho(x)\Delta(x)^p\,\dd x.
 \label{eq:supp_ring_small_current_mass}
\end{align}

For large currents, Eq.~\eqref{eq:supp_ring_current} and
the Cauchy--Schwarz inequality imply that
\begin{align}
 L^2
 &\leq
 M_p(\jmath)\int_0^L\frac{\dd x}{\rho(x)z_{\jmath}(x)^p}
 \notag\\
 &=M_p(\jmath)\frac{LF}{\jmath}
 \label{eq:supp_ring_mass_lower_bound}
\end{align}
holds. Thus, as $\jmath\to\infty$,
$M_p(\jmath)\geq L\jmath/F\to\infty$.

\redtext{Since the periodic solution $z_{\jmath}$ depends continuously on
$\jmath$ and is strictly increasing in $\jmath$ at every point, $M_p(\jmath)$
is continuous and strictly increasing.}

\redtext{Consequently, by continuity and strict monotonicity, the
normalization condition $M_p(\jmath)=1$ has a unique solution $\jmath>0$
if and only if
\begin{align}
 \int_0^L\rho(x)\Delta(x)^p\,\dd x
 &<1
 \label{eq:supp_ring_criterion}
\end{align}
holds.
Thus, condition~\eqref{eq:ring_example_solvability} is necessary and
sufficient for Eqs.~\eqref{eq:supp_ring_bvp_equation}--\eqref{eq:supp_ring_bvp%
_normalization} to have a positive normalized periodic solution, and $\gamma$
and $\jmath$ are unique whenever a solution exists.}

\subsection{Verification of solvability in the numerical example}

Using the same initial density and parameters as in the main text,
we considered all output times in $0\leq t\leq3$,
used Eq.~\eqref{eq:ring_example_delta} to compute $\Delta_t(x)$,
and directly evaluated the left-hand side of
Eq.~\eqref{eq:supp_ring_criterion}.

In both cases, the maximum was attained at the initial time. The values
\begin{align}
 \max_{\substack{0\leq t\leq3\\\text{output times}}}
 \int_0^L\rho_t(x)\Delta_t(x)^p\,\dd x
 &\simeq
 \begin{cases}
 0.4220,&p=1,\\
 0.5941,&p=2
 \end{cases}
 <1
 \label{eq:supp_ring_uniform_solvability}
\end{align}
show that, at all output times examined,
both reverse KL and Hellinger satisfy the necessary and sufficient condition
for solvability,
numerically confirming the existence and uniqueness
of a positive normalized relaxation target.
\end{appendix}

\end{document}